\documentclass[journal,10pt]{IEEEtran}
\usepackage{graphicx}
\usepackage{epstopdf}
\usepackage{epsfig}
\usepackage{caption}
\usepackage{subcaption}
\usepackage{mathtools}
\usepackage{bm,amsmath, amssymb}
\usepackage{amsmath}
\usepackage{amsfonts}
\usepackage {amssymb,amsmath}
\usepackage{cite}
\usepackage{epstopdf}
\usepackage{color}
\usepackage{microtype}
\usepackage{float}
\usepackage[noend]{algpseudocode}
\usepackage{algorithmicx,algorithm}
\usepackage{booktabs}
\usepackage{threeparttable}
\usepackage{stfloats}

\DeclareGraphicsExtensions{.pdf,.png,.jpg}

\usepackage{cite}

\begin{document}
	\title{\vspace{-8mm} Orthogonal Chirp Delay-Doppler Division Multiplexing (CDDM) Modulation for High Mobility Communications}
	\author{\IEEEauthorblockN{Chaoyuan Bai, Pingzhi Fan, Zhengchun Zhou, Zilong Liu}
		\\
		
		\thanks{			
			C. Bai, P. Fan, and Z. Zhou are Information Coding \& Transmission Key Lab of Sichuan Province, Southwest Jiaotong University, Chengdu, 611756, China. (e-mail: swjtubcy@my.swjtu.edu.cn, pzfan@swjtu.edu.cn, zzc@swjtu.edu.cn)
			
			Z. Liu is with the School of Computer Science and Electronic Engineering, University of Essex, UK, (e-mail:zilong.liu@essex.ac.uk)}
	}

	\maketitle
	
	\vspace{-25mm}
	\begin{abstract}  
		This paper proposes a novel multi-carrier modulation framework for high-mobility communication scenarios. Our key idea lies in spreading data symbols across the delay-Doppler (DD) domain through orthogonal chirp-Zak transform (CZT). To enable efficient signal multiplexing, the proposed modulation scheme employs a transmitter signal that maintains orthogonality with the inherent resolution characteristics of the DD plane. Termed as Orthogonal Chirp Delay-Doppler Division Multiplexing (CDDM), we demonstrate a synergistic integration of chirp waveform properties with the channel structure of the DD domain, thereby achieving advantages with both lower computational efficiency and improved detection performance. We introduce a novel CZT-based superimposed sparse pilot structure to enable simultaneous estimation of delay-Doppler shifts and channel coefficients. For enhanced performance, we further develop an embedded pilot scheme that demonstrates channel estimation performance comparable to that of Orthogonal Delay-Doppler Division Multiplexing (ODDM) systems. Simulation results demonstrate that CDDM achieves significant bit error rate (BER) improvements over existing modulation schemes , under perfect channel state information (CSI), as well as superior out-of-band emissions (OOBE). Further, for the imperfect CSI case, the proposed CZT-based superimposed pilot scheme leads to significantly reduced normalized mean square error (NMSE), whilst attaining equivalent estimation accuracy to that of ODDM with lower computational complexity. 
		
	\end{abstract} 
	
	\vspace{-2mm}
	\normalsize
	\section{Introduction}
	\label{secintro}
	\subsection{Background}
	High-mobility communications play an important role in the upcoming sixth generation (6G) mobile systems \cite{10152009}. In 6G, a challenging target is to provide reliable communications for moving speeds of 1000 km/h or higher in various dynamic environments (e.g., high-speed railways, vehicle-to-everything networks, flying drones, low-earth-orbit (LEO) satellite systems). The legacy Orthogonal Frequency Division Multiplexing (OFDM), while widely adopted, may suffer from poor error rate performance in high-mobility channels due to the orthogonality loss caused by large Doppler \cite{10012421}.
	
	Chirp signals \cite{7523229}, characterized by a linearly varying instantaneous frequency over time, exhibit inherent robustness against severe channel impairments commonly encountered in high-mobility wireless environments. First, their excellent Doppler resilience ensures that the signal energy remains well-localized in the delay-Doppler domain, enabling reliable detection even under high-speed movement. Second, chirp signals enjoy low or zero autocorrelation sidelobes, which are helpful for mitigating inter-symbol interference caused by multipath reflections. Third, their inherent time-frequency diversity facilitates efficient exploitation of doubly selective channels, thus leading to improved link reliability. These properties make chirp waveforms particularly attractive for emerging applications such as vehicular-to-everything (V2X) communication, UAV swarms, and high-speed rail systems, where traditional orthogonal waveforms struggle to maintain good performance. In Orthogonal Chirp Division Multiplexing (OCDM) \cite{7523229}, for example, data symbols are modulated using mutually orthogonal chirps via the Discrete Fresnel Transform (DFnT). More recently, Chirp-Convolved Data Transmission (CCDT) \cite{9284570} and Affine Frequency Division Multiplexing (AFDM) \cite{9562168} have been proposed as two representative chirp-based modulation schemes. CCDT allows one to adjust the coefficients of the quadratic term to achieve improved ambiguity function, thereby meeting diverse velocity and distance detection requirements of various scenarios. AFDM, in contrast, selects the coefficients of the quadratic term coupled to the maximum Doppler shift and is capable of achieving full diversity in doubly selective channels.
	
	Besides, high-mobility communications can also be achieved in the Delay-Doppler (DD) domain. A prominent example is Orthogonal Time Frequency Space (OTFS) \cite{7925924}, which offers excellent channel diversity and relatively static channel state information (CSI) \cite{10279816}. However, OTFS may not be backward-compatible with OFDM as it requires a two-dimensional DD transform. Moreover, ideal OTFS modulation relies on pulses that satisfy the biorthogonal robustness condition, which is unattainable due to the uncertainty principle. Rectangular pulses are commonly employed in OTFS, often resulting in significant out-of-band emissions (OOBE).
	
	To address the challenge of designing realizable orthogonal pulses for the DD plane, Orthogonal Delay-Doppler Division Multiplexing (ODDM) was introduced in \cite{9829188}. ODDM is a staggered, upsampled-OFDM-based modulation scheme that enables multi-carrier modulation in the DD domain, thereby enhancing signal alignment with DD channels. The transmission pulse proposed in ODDM, referred to as the Delay-Doppler Orthogonal Pulse (DDOP), is designed as a pulse train that preserves orthogonality with respect to the resolution of the DD plane. In comparison, ODDM exhibits superior performance compared to OTFS in terms of both OOBE and bit error rate (BER).
	
	\subsection{Related Works}
	A plethora of waveform studies have been conducted, aiming to achieve reliable communications in high mobility scenarios.
	
	Similar to OFDM, it was shown in \cite{7523229} that chirp-based waveforms may suffer from poor OOBE performance. Such an issue can be mitigated by windowing techniques, as pointed out in \cite{10490647}. By properly tuning the chirp rate, AFDM generally outperforms OCDM, where the latter may only achieve partial diversity \cite{10490647}, \cite{9743516}. Compared to OTFS, AFDM enjoys a lower training overhead whilst achieving comparable BER performance \cite{9746329}, thanks to its one-dimensional signal generation structure. For improved backward compatibility with OFDM systems, \cite{10580928} demonstrated that AFDM can be efficiently implemented as a Discrete Fourier Transform (DFT) precoded OFDM scheme. The integration of AFDM with sparse code multiple access (SCMA), called AFDM-SCMA, was proposed in \cite{10566604} to support massive connectivity over high mobility channels. Utilizing the redundant information in the chirp periodic prefix of AFDM symbols, synchronization can be effectively implemented for AFDM using maximum-likelihood criteria \cite{10571184}.

	DD domain waveforms are attractive due to efficient channel estimation \cite{9632684}, full-diversity data transmission \cite{8892482}, excellent coded system performance \cite{9404861}, and low-complexity precoding for MIMO transmissions \cite{10042436}. \cite{shtaiwi2024orthogonal} demonstrated its unique advantages for integrated sensing and communication (ISAC) systems, while \cite{10584089} provided crucial insights into MIMO-OTFS implementations with non-orthogonal multiple access (NOMA).  \cite{10043628} showed its effectiveness for LEO satellite links, whereas \cite{8085125} suggested OTFS as a leading waveform candidate for future networks. OTFS also inspired Orthogonal Time Sequency Multiplexing (OTSM) as reported in \cite{9417451}. \cite{10579490} discussed the relationship of DD domain modulation schemes, attempting to establish an unifying framework by introducing the transformation relations between different domains. New pilot sequences can be developed by exploiting the representation of orthogonal chirp in the DD domain for more accurate channel estimation \cite{10772237}.

	\subsection{Motivations and Contributions}
	Despite numerous research attempts in DD domain waveforms, the existing state-of-the-art researches have not synergistically exploited the advantages of DD channel by fully taking advantage of the properties of chirp signal. Although \cite{10772237} utilized the chirp in the DD domain as pilot sequence, the channel estimation at the receiver suffers from significantly increased complexity by fully taking advantage of the properties. Inspired by this observation, this paper proposes to transmit data symbols via chirp signals in the DD domain. It is found that orthogonal chirp signals can be transformed into DD domain with partial orthogonality that facilitates data separation in DD plane. It is also shown that pulse-shaped DD domain multicarrier modulation can simultaneously preserve the inherent advantages of chirp waveforms while effectively suppressing OOBE. At the receiver, after multicarrier demodulation, the sparse structure of the DD channel enables rapid data processing by exploiting the correlation of DD-domain chirps. Therefore, our proposed scheme can fully utilize the DD channel properties and chirp signal advantages, whilst achieving significant OOBE reduction.

	
	The main contributions of this paper are as follows:
	
	\begin{itemize}
		\item The Chirp Zak Transform (CZT) and its inverse are proposed to efficiently transform data symbols into the DD domain by spreading over DD chirps, thus enhancing the signal detection in high mobility scenarios.
		\item Building on CZT, orthogonal chirp delay-Doppler division multiplexing (CDDM) is proposed. It is shown that the proposed CDDM leads to reduced OOBE via pulse shaping, and enables efficient DD-domain channel estimation.
		\item 
		By exploiting the correlation of DD chirps, we present novel CDDM channel estimation and data detection schemes to enable accurate channel estimation and reliable data recovery.
		\item A superimposed pilot scheme and an embedded pilot scheme are designed for CDDM. The superimposed pilot scheme, based on IDAFT, ensures fast and accurate channel estimation using threshold. The embedded pilot scheme, implemented within the CDDM system based on CZT, provides more precise estimation.
		\item The BER of data recovery and the normalized mean square error (NMSE) of channel estimation for CDDM are evaluated through simulations and compared against several existing waveforms. It is shown that the proposed CDDM demonstrates superior performance in both BER and NMSE, with more effective superimposed pilots and comparable embedded pilot performance.
		
	\end{itemize}
	
	\subsection{Organization of This Work}
	The subsequent sections are organized as follows: Section II introduces the CZT, Section III presents the CDDM modulation and demodulation scheme, Section IV discusses the DD domain input and output of CDDM, Section V covers the channel estimation and data detection methods for CDDM, Section VI introduces the new superimposed pilot scheme and embedded pilot scheme for CDDM, Section VII provides a performance analysis comparison, and Section VIII concludes the paper.
	
	\section{Chirp-Zak Transform} 	\label{CZT}
	
		A chirp signal is a waveform with time-varying frequency, which can be transformed into its DD domain representation via the Zak transform. Denote by $\Delta f$ and $T$ the subcarrier spacing and the symbol duration, respectively, let us consider a DD plane $\Gamma = [\frac{m}{M_D \Delta f}, \frac{n}{N_D T}]=[\frac{mT}{M_D} , \frac{n}{N_D T}] $ for $ m=0,1,...,M_D -1$ and $ n = 0,1,...,N_D -1 $, with $M_D$ representing the number of delay indices and $N_D$ representing the number of Doppler indices. Denote by $N=M_D N_D$ the total number of DD grids. Further, assume that $ M_D = L N_D$, where $L$ is an integer. One can see that $ \frac{T}{M_D} $ represents the delay resolution and $ \frac{1}{N_D T} $ represents the Doppler resolution. Consider $N$ data symbols $\bm x = [x(0), x(1), ..., x(i), ..., x(N-1)]$. 
		
		Let $\varphi_i (w)$ denote to the $i$-th discrete orthogonal chirp of $x(i)$, which is formally defined as $\varphi_i (w) = e^{j \frac{\pi}{4}} e^{-j \pi \frac{(w-i)^2}{N}}$, with $w = 0,1,...,N-1$. 
	\begin{figure}[h]
		\centering
		\includegraphics[width = 220px]{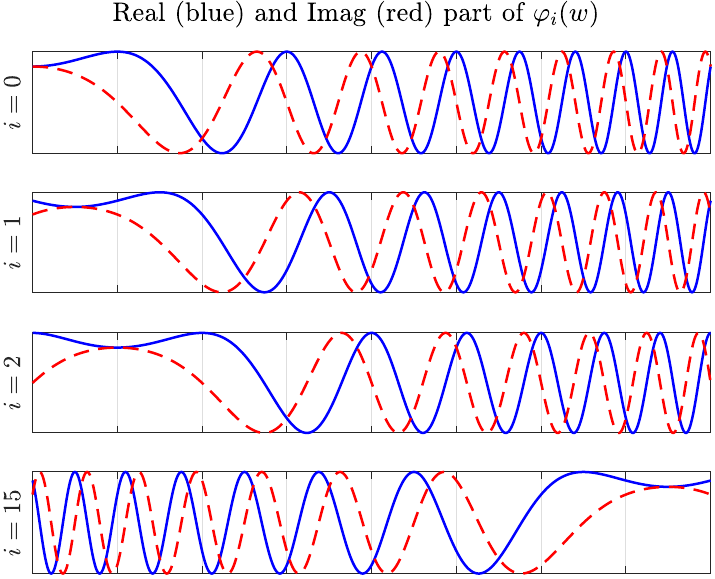}
		\caption{The real and imaginary parts of $\varphi_i(w)$ when $N = 16$.}
		\label{realandimagpartofocdm}
	\end{figure}
	
	Fig. \ref{realandimagpartofocdm} demonstrates that $\varphi_i (w)$ are mutually orthogonal in the chirp dimension, and both the amplitude and phase of each chirp can be exploited for modulation through the Inverse DFnT (IDFnT) \cite{7523229}. Moreover, the linearly increasing frequency of chirp signals enhances their robustness in complex transmission channels, while their favorable perfect correlation properties further facilitate efficient data processing.

	By applying the IDFnT to $\bm x$, the discrete orthogonal chirp signals $s_c(q)$ can be obtained
	\begin{eqnarray}
		\label{ocdm}
		\begin{split} 
			s_c(q)= \sum_{w=0}^{N-1} x(w) \varphi_w (q), \\
			q = 0,1,...,N-1.
		\end{split}
	\end{eqnarray}
	
	Applying the discrete Zak transform (DZT) to $s_c$ yields its DD domain representation
	\begin{eqnarray}
		\label{zakocdm}
		\begin{split} 
			&\mathcal{Z}_{s_c} [m,n] = \frac{1}{\sqrt{N_D}} \sum_{k=0}^{N_D -1} s_c(m+k M_D) e^{-j 2 \pi \frac{nk}{N_D}}\\
			&= \frac{1}{\sqrt{N_D}} \sum_{k=0}^{N_D -1} \sum_{w=0}^{N-1} x(w) \varphi_w (m+k M_D) e^{-j 2 \pi \frac{nk}{N_D}}.
		\end{split}
	\end{eqnarray}
	
\begin{figure}[h]
		\centering
		\includegraphics[width = 220px]{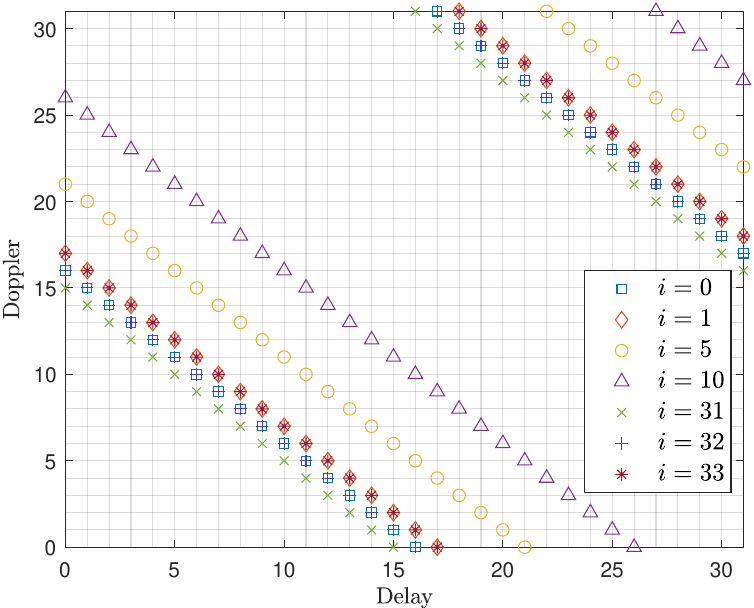}
		\caption{The distribution of $\sum_{i}^{ }\mathcal{CZ}_{x(i)}$ when $M_D = N_D = 32$ and $i=0,1,5,10,31,32,33$.}
		\label{CZTOCDM}
	\end{figure}

	In this context, the DD domain data symbols can be viewed as the superposition of individually precoded data symbols, where the precoding matrix is the DD domain representation of orthogonal chirp signals. We define this process as the CZT of $\bm x$, which transform $\bm x$ into DD domain data symbols while spreading them across the DD chirp. Then, the CZT coefficient of the $i$-th data symbol $x(i)$ can be defined as
	\begin{eqnarray}
		\label{cztforxi}
		\begin{split} 
			\mathcal{CZ}_{x(i)} [m,n]&= \frac{x(i)}{\sqrt{N_D}} \sum_{k=0}^{N_D -1} \varphi_i (m+k M_D) e^{-j 2 \pi \frac{nk}{N_D}}\\
			&= \frac{x(i)}{\sqrt{N_D}} \sum_{k=0}^{N_D -1} \varphi_i (m) e^{-j2 \pi \frac{ kf_i }{N_D}} \\
			&= \begin{cases} \sqrt{N_D}x(i)\varphi_i (m) &,{\text{if} \  [ f_i ]_{N_D} = 0}\\ 0 &, \text{otherwise}\end{cases},
		\end{split}
	\end{eqnarray}
	where $f_i =\frac{M_D}{2} + m+n-i$, $[\cdot]_{N_D}$ stands for the mod $N_D$ operator. Therefore, the CZT of $\bm x$ is $\sum_{i=0}^{N-1}\mathcal{CZ}_{x(i)}$. In this case, $x(i)$ can be recovered by the following Inverse CZT (ICZT),
	\begin{eqnarray}
		\begin{split} 
			x(i) = \frac{1}{N\sqrt{N_D}} \sum_{m=0}^{M_D-1} \sum_{k=0}^{N_D-1} \sum_{n=0}^{N_D-1} \mathcal{CZ}_{x(i)} [m,n] \\ e^{j 2\pi \frac{nk}{N_D}} \varphi_i^* (m+kM_D),
		\end{split}
	\end{eqnarray}
	where $\varphi_i^*(\cdot)$ represents the conjugate of $\varphi_i(\cdot)$. One can readily show that $\mathcal{CZ}_{x(i)}$ is a sparse matrix, whereby each row contains one non-zero entry only. For example, when $M_D = N_D = 32$, $\sum_{i}^{ }\mathcal{CZ}_{x(i)}$ is shown in Fig. \ref{CZTOCDM}, where $i=0,1,5,10,31,32,33$.
	From (3), it can be observed that when $M_D$ and $N_D$ are fixed, the non-zero entries in $\mathcal{CZ}_{x(i)}$ remain unchanged for given $i$, and thus can be computed in advance. Based on this, the fast implementation of CZT has been present, and the CZT for $x(i)$ can be rewritten as
	\begin{eqnarray}
		\begin{split}
			\mathcal{CZ}_{x(i)} &= x(i)\bm{\mathrm \Phi}_{i,DD},
		\end{split}
		\label{cztfast1}
	\end{eqnarray}
	where $\bm{\mathrm \Phi}_{i,DD} = \frac{1}{\sqrt{N_D}}\bm{\mathrm \Phi}_{i} \bm{\mathrm F_{N_D}}$, $\bm{\mathrm F_{N_D}}$ is $N_D \times N_D$ DFT matrix, $\bm{\mathrm \Phi}_{i}$ is defined as
	\begin{eqnarray}
		\begin{split}
			&\bm{\mathrm \Phi}  _i =  \\&
			{
				\begin{bmatrix}		
					\varphi _ i (0) & \varphi _ i (M_D) &  ... & \varphi _ i ((N_D-1)M_D)  \\
					\varphi _ i (1) & \varphi _ i (M_D + 1) & ... & ...\\
					...& ... &...& ...\\
					\varphi _ i (M_D-1) & \varphi _ i (2M_D-1)  & ... & \varphi _ i (N_D M_D -1)
				\end{bmatrix}
			}.
		\end{split}
	\end{eqnarray}
	
	For example, when $M_D = N_D = 6$, the $\bm{\mathrm \Phi}_{0,DD}$ is given by
	\begin{eqnarray}
		\begin{split}
			\label{example1}
			\bm{\mathrm \Phi}_{0,DD} = 
			\sqrt{6}	
			\small{
				\setlength{\arraycolsep}{3pt} 		
				\begin{bmatrix}			  
					0& 0& 0& \varphi_0(0)& 0& 0\\
					0& 0& \varphi_0(1)& 0& 0& 0\\
					0& \varphi_0(2)& 0& 0& 0& 0\\
					\varphi_0(3)& 0& 0& 0& 0& 0\\
					0& 0& 0& 0& 0& \varphi_0(4)\\
					0& 0& 0& 0& \varphi_0(5)& 0\\
				\end{bmatrix},
			}
		\end{split}
	\end{eqnarray}
	
	Clearly, when the DD domain dimension is determined, the matrix $\bm{\mathrm \Phi}_{i,DD}$ does not contain random elements, and it can be precomputed. On the other hand, the CZT for $\bm x$ requires only summing $N$ sparse matrices with the complexity of $\mathcal{O}(2M_DN)$ when Compressed Sparse Row (CSR) format is used to store sparse matrices \cite{8849306}.
	
	Performing the ICZT to recover $\bm x$ from $\mathcal{CZ}_{\bm x}$ requires Inverse DZT (IDZT) followed by an IDFnT, leading to a computational complexity of $\mathcal{O}(N\log N+2N+N\log N_D)$ which may not be affordable for practical implementations.
	
	From \eqref{cztforxi}, it is evident that after the CZT, the non-zero entries of DD-domain chirps corresponding to $x(i)$ and $x(i + cN_D)$, where $c \in \{0,1,\ldots,M_D-1\}$, share identical distributions. At the same time, they exhibit a form of orthogonality. In contrast, the same cannot be achieved by non-overlapping symbols. This property facilitates a fast ICZT for data symbol recovery.
	
	Since the CZT matrix can be precomputed, one can determine the DD domain positions of each data symbol’s non-zero entries. Define
	\begin{equation}
		\Delta_i = \{(m'_{i,0}, n'_{i,0}), (m'_{i,1}, n'_{i,1}), \ldots, (m'_{i,M_D-1}, n'_{i,M_D-1})\}
		\label{positiondelta}
	\end{equation}
	denote the set of positions corresponding to non-zero entries of $x(i)$ after the CZT, where $(m'_{i,m}, n'_{i,m})$ represents the position at the $m$-th delay index. For instance, in \eqref{example1}, we have
	\begin{equation}
		\Delta_0 = \{(0,3), (1,2), (2,1), (3,0), (4,5), (5,4)\}.
		\label{delta0}
	\end{equation}
	
	Let $\Delta = \{\Delta_0, \Delta_1, \ldots, \Delta_{N-1}\}$ be the distribution set for all data symbols. Then, one can achieve fast implementation of ICZT by
	\begin{equation}
		x(i) = \frac{1}{M_D \sqrt{N_D}} \sum_{m=0}^{M_D-1} \mathcal{CZ}_{\bm x,i}(m) \, \varphi_i^*(m),
		\label{FICZT}
	\end{equation}
	where $\mathcal{CZ}_{\bm x,i}(m) = \mathcal{CZ}_{\bm x}[m'_{i,m}, n'_{i,m}]$ denotes the non-zero entry associated with $x(i)$ under the $m$-th delay index, retrievable via $\Delta_i$. This approach reduces the complexity of recovering $\bm{x}$ to $\mathcal{O}(2M_D N)$ by restricting computation to symbol-relevant non-zero entries only.

	\section{CDDM Modulation} 	\label{CDDM Modulation}
\begin{figure*}[h]
		\centering
		\includegraphics[width = .95\textwidth]{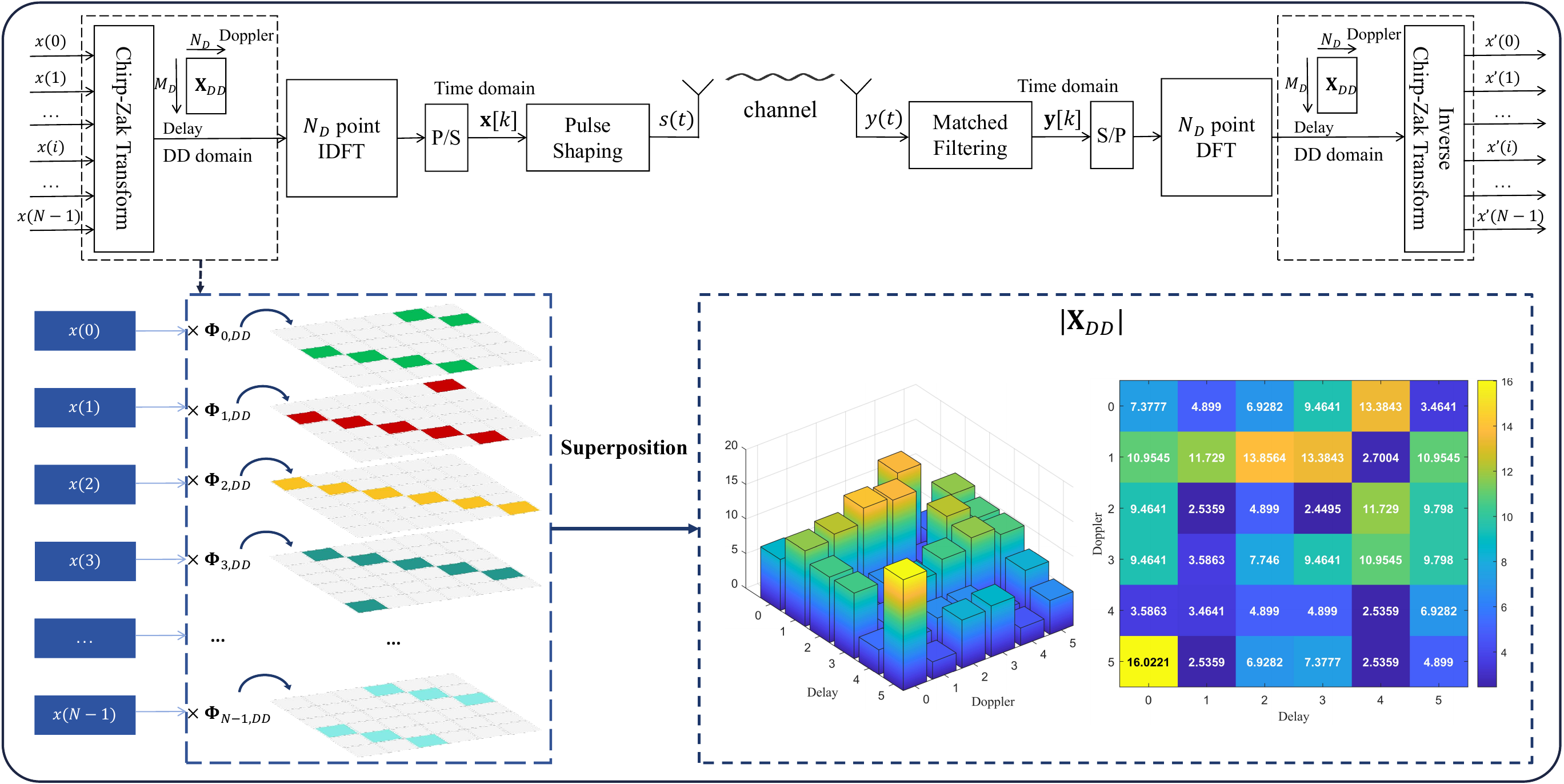}
		\caption{CDDM system model. In $\bm{\mathrm X}_{DD}$, the value within each grid is the superposition of multiple random complex values. As a result, random constructive or destructive interference may occur, leading to large differences in the values within each grid.}
		\label{system model}
	\end{figure*}

	Let us assume that there are $N$ data symbols which are grouped to be a vector (e.g., via QPSK), $\bm x = [x(0),x(1),...,x(i),...,x(N-1)]$. In our proposed CDDM modulation, as shown in Fig. \ref{system model}, the data symbols $\bm x$ is firstly transformed into DD domain $\bm{\mathrm X}_{DD}$ by CZT in \eqref{cztfast1} as follows:	
	\begin{eqnarray}
		\label{DDdataSymbolofTx}	
		\begin{split}
			\bm{\mathrm X}_{DD}[m,n] &= \sum_{i=0}^{N-1} \sum_{k=0}^{N_D -1}\frac{x(i)}{\sqrt{N_D}} \varphi_i (m+k M_D) e^{-j 2 \pi \frac{nk}{N_D}}\\
			&=\sum_{i=0}^{N-1} x(i)\bm{\mathrm \Phi}_{i,DD}[m,n].	
		\end{split}
	\end{eqnarray}
	
	\begin{figure}[h]
		\centering
		\includegraphics[width = 220px]{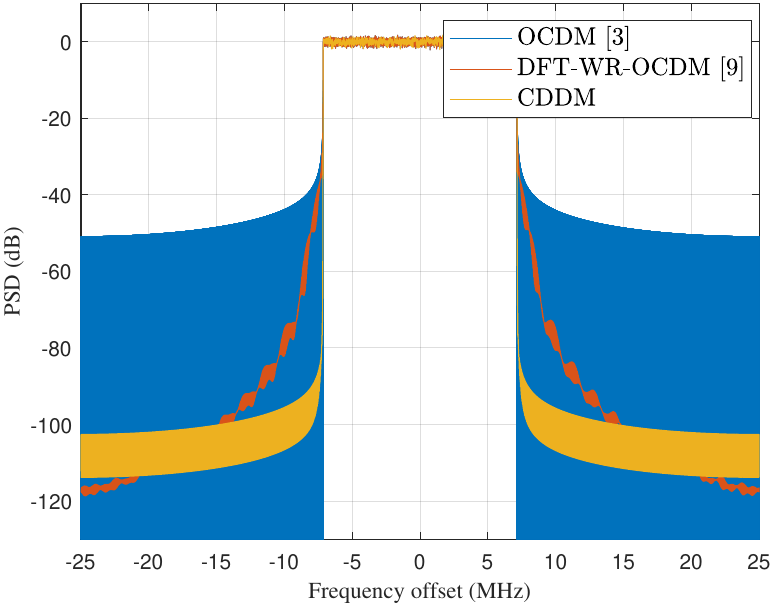}
		\caption{PSD comparison, $M_D = 512, N_D = 32$, QPSK. }
		\label{OOBEPSD}
	\end{figure}
	
	As shown in the Fig. \ref{system model}, CZT in the CDDM system can be viewed as pre-coding for the DD domain data symbols. $\bm{\mathrm x}[k]$ in Fig. \ref{system model} represents the time domain discrete transmission sequence which is obtained via IDFT as follows: 
	\begin{eqnarray}
		\begin{split}
			\bm{\mathrm x} [k] = &\frac{1}{\sqrt{N_D}} \sum_{n=0}^{N_D-1} \bm{\mathrm X}_{DD}[[k]_{M_D}, \lfloor k/M_D \rfloor] e^{j2\pi \frac{\lfloor k/M_D \rfloor}{N_D}n},\\
			&k = 0,1,...,M_DN_D-1,
		\end{split}
	\end{eqnarray}
	where $\lfloor \cdot \rfloor$ represents the floor operation. At the receiver, after the multi-carrier demodulation, ICZT is applied to convert DD domain data symbols to $\bm x' = [x'(0),x'(1),...,x'(i),...,x'(N-1)]$ for data detection, whereby $x'(i)$ is the $i$-th recovered data symbol. Denote by $i_{m,n}$ the index of the data symbols located in the grid $[m,n]$. It follows from \eqref{cztforxi} that
	\begin{eqnarray}
		\label{indicesingrids}
		\begin{split}
			i_{m,n} = \left[ m+n+ \frac{M_D}{2} \right]_{N_D} + cN_D, \\
			c = 0,1,...,M_D-1.
		\end{split}
	\end{eqnarray}

	Based on \eqref{cztforxi} and \eqref{indicesingrids}, $\bm{\mathrm X}_{DD}$ can be defined as
	\begin{eqnarray}
		\begin{split}
			\bm{\mathrm X}_{DD}[m,n] = \sum_{c=0}^{M_D-1} \sqrt{N_D}x(i_{m,n})\varphi_{i_{m,n}}(m).
		\end{split}
	\end{eqnarray}
	
	Suppose that $g(t)$ is a DD domain pulse which is orthogonal according to the DD resolutions \cite{9829188}. Let us write $g(t)$ as follows: 
	\begin{eqnarray}
		\label{eqn_16}
		\begin{split}
			g(t) = \sum_{n=0}^{N_D-1} a(t-nT),
		\end{split}
	\end{eqnarray}
	where $a(t)$ stands for a time-symmetric real-valued square-root Nyquist pulse, satisfying $\int_{-\infty}^{\infty}|a(t)|^2 dt= \frac{1}{N_D}$. Then, the CDDM waveform can be expressed as:
	\begin{eqnarray}
		\label{st}
		\begin{split}
			s(t) =\sqrt{N_D} \sum_{m=0}^{M_D-1} &\sum_{n=0}^{N_D-1}\sum_{c=0}^{M_D-1} x(i_{m,n})\varphi_{i_{m,n}}(m) \\
			&g\left(t-\frac{mT}{M_D}\right) e^{j 2 \pi \frac{n}{N_DT}\left(t-\frac{mT}{M_D}\right)}.
		\end{split}
	\end{eqnarray}

	In the proposed CDDM scheme, each data symbol is mapped to the DD domain through DD chirp spreading. Despite the superposition of different data symbols in the DD domain, they remain effectively distinguishable due to the correlation of the DD chirps. At the receiver, multi-carrier demodulation is performed first, followed by the estimation of the delay and Doppler shift for each path based on the sparsity of the DD channel. Subsequently, the path gains are accurately estimated by exploiting the correlation properties of the DD chirps, enabling reliable data recovery.

	The DFT employed in the CZT and the IDFT used to generate the time domain samples result in a waveform that resembles an orthogonal chirp waveform during transmission. As a consequence, CDDM can be viewed as a filtered orthogonal chirp signal when transmitting, with the filter being a Nyquist square-root pulse, typically a square-root raised cosine (SRRC) pulse. Unlike conventional orthogonal chirp waveforms, CDDM not only retains the robust transmission characteristics of chirp signals but also leverages the SRRC pulse to effectively suppress the OOBE of the chirp signal, as illustrated in Fig. \ref{OOBEPSD}. The power spectral density (PSD) of OCDM, DFT-WR-OCDM, and CDDM are compared in Fig. \ref{OOBEPSD}. Benefiting from SRRC pulse shaping, CDDM achieves reduced OOBE, its OOBE performance is comparable to that of DFT-WR-OCDM in \cite{10490647}. Additionally, the SRRC pulse significantly reduces interference in the DD domain, improving the accuracy of correlation computations. Thus, CDDM integrates the advantages of chirp signals and DD channel. 
	
	\section{Input-Output relation of CDDM} 	\label{IO of CDDM}
	
	Suppose that a doubly-selective channel is composed of $P$ pathwhich can be expressed as
	\begin{eqnarray}
		\begin{split}
			h(\tau,v) = \sum_{p=1}^{P} h_p \delta(\tau-\tau_p)\delta(v-v_p),
		\end{split}
	\end{eqnarray}
	where $h_p, \tau_p, v_p$ denote the channel gain, delay, and Doppler of the $p$-th path, respectively.  Due to the band-pass filtering and sampling, the observed $h(\tau,v)$ is usually modeled by a discrete equivalent channel with
	\begin{eqnarray}
		\begin{split}
			\tau_p = \frac{l_pT}{M_D}, v_p = \frac{k_p}{N_DT}, 
		\end{split}
	\end{eqnarray}
	where $l_p$ and $k_p$ represent the delay and Doppler indices in the DD domain for the $p$-th path, respectively.
	
	From \eqref{st}, the received CDDM signal at the receiver is  
	
	\begin{eqnarray}
		\label{yt}
		\begin{split}
			y(t)&=\sum_{p=1}^{P} h_ps(t-\tau_p)e^{j2\pi v_p(t-\tau_p)}+z(t),\\
			&=\sum_{p=1}^{P}\sum_{m=0}^{M_D-1}\sum_{n=0}^{N_D-1} \sum_{c=0}^{M_D-1} h_p x(i_{m,n})\varphi_{i_{m,n}}(m) \\ 
			&g\left(t-\frac{mT}{M_D}-\tau_p\right) e^{j2\pi \left( \frac{n+k_p}{N_DT}\left(t-(m+l_p)\frac{T}{M_D}\right) \right)} \\ 
			&e^{j2\pi \frac{k_p}{M_DN_D}}+z(t).
		\end{split}
	\end{eqnarray}
	
	After the matched filtering using $g(t)$, the signal in DD domain can be obtained, the signal at the $n$-th subcarrier of the $m$-th CDDM symbol is
	\begin{eqnarray}
		\label{ydd}
		\begin{split}
			\bm{\mathrm Y}_{DD}[m,n] = \sum_{p=1}^{P} h_p \bm{\mathrm{\widehat{X}}}_{DD}[\widehat{m},\widehat{n}] e^{j2\pi \frac{k_p(m-l_p)}{M_DN_D}}+z[m,n],
		\end{split}
	\end{eqnarray}
	where $\widehat{m} = m-l_p$, $\widehat{n} = [n-k_p]_{N_D}$, $\bm{\mathrm{\widehat{X}}}_{DD}[\widehat{m},\widehat{n}] = \bm{\mathrm{X}}_{DD}[\widehat{m},\widehat{n}]$ for $\widehat{m} \ge 0$, and $\bm{\mathrm{\widehat{X}}}_{DD}[\widehat{m},\widehat{n}] = \bm{\mathrm{X}}_{DD}[M_D+\widehat{m},\widehat{n}]e^{-j2\pi\frac{\widehat{n}}{N_D}}$ for $\widehat{m} < 0$. Without loss of generality, lets define the sets of normalized delay and Doppler shifts by $\mathcal{L}=\{l_1,l_2,...,l_P\}$ and $\mathcal{K}=\{k_1,k_2,...,k_P\}$.  Here, this paper assume that each path only possesses integer delay and Doppler shifts. Then, the DD channel can be written as
	\begin{eqnarray}
		\label{DDchannel}
		\begin{split}
			h[l,k] = \begin{cases} h_p &,{\text{if} \  l=l_p,k=k_p}\\ 0 &, \text{otherwise}\end{cases},
		\end{split}
	\end{eqnarray}
	where $l \in \mathcal{L}$ and $k \in \mathcal{K}$. To facilitate understanding, it can be expressed in matrix form. Let $\bm{\mathrm{x}}_m$, $\bm{\mathrm{y}}_m$ be the transmitted and received symbol vectors at the $m$-th delay index. Then, $\bm{\mathrm{X}}_{DD}$ and $\bm{\mathrm{Y}}_{DD}$ can be vectorized as $\bm{\mathrm{x}}_{DD} =\{\bm{\mathrm{x}}_0^T, \bm{\mathrm{x}}_1^T,...,\bm{\mathrm{x}}_{M_D-1}^T\}^T$, $\bm{\mathrm{y}}_{DD} =\{\bm{\mathrm{y}}_0^T, \bm{\mathrm{y}}_1^T,...,\bm{\mathrm{y}}_{M_D-1}^T\}^T$. Similarly, the noise matrix is $\bm{\mathrm{z}}_{DD} =\{\bm{\mathrm{z}}_0^T, \bm{\mathrm{z}}_1^T,...,\bm{\mathrm{z}}_{M_D-1}^T\}^T$. After matched filtering, the DD domain input-output relation of the CDDM system can be expressed as a linear system
	\begin{eqnarray}
		\label{DDIOsystem}
		\begin{split}
			\bm{\mathrm{y}}_{DD}=\bm{\mathrm{H}}\bm{\mathrm{x}}_{DD}+\bm{\mathrm{z}}_{DD},
		\end{split}
	\end{eqnarray}
	where $\bm{\mathrm{H}} \in \mathbb{C}^{M_DN_D \times M_DN_D}$ is the effective DD channel matrix which is combined of sub-channel matrices $\bm{\mathrm{H}}_{m,l} \in \mathbb{C}^{N_D \times N_D}$. The sub-channel matrices can be expressed as
	\begin{eqnarray}
		\label{sub-channelmatrix}
		\begin{split}
			\bm{\mathrm{H}}_{m,l} = \begin{cases} \sum_{k \in \mathcal{K}} h[l,k]e^{j2\pi \frac{k(m-l)}{M_DN_D}}\bm{\mathrm{C}}^k  &,{if \  \hat{m} \geq 0}\\ \sum_{k \in \mathcal{K}} h[l,k]e^{j2\pi \frac{k(m-l)}{M_DN_D}}\bm{\mathrm{C}}^k\bm{\mathrm{D}} &, {if \  \hat{m} < 0}\end{cases},
		\end{split}
	\end{eqnarray}		
	where $\bm{\mathrm{C}}$ is an $N_D\times N_D$ cyclic permutation matrix. 

Due to the CP, there is an additional phase rotation when $\hat{m} < 0$. For the $n$-th subcarrier, the additional phase rotation is $e^{-j2\pi\frac{n}{N_D}}$. Then, this paper define $\bm{\mathrm{D}}$ which refers to the $N_D\times N_D$ additional phase rotation matrix below: 
\begin{eqnarray}
	\label{phasematrix}
	\begin{split}
		\bm{\mathrm{D}} = 
		\begin{bmatrix}			  
			1&0&\cdots&0\\
			0&e^{-j2\pi\frac{1}{N_D}}&\cdots&0\\
			\vdots&&\ddots&\vdots\\
			0&\cdots&0&e^{-j2\pi\frac{N_D-1}{N_D}}\\
		\end{bmatrix}.
	\end{split}
\end{eqnarray}

Let $\mathfrak{L} = max\{\mathcal{L}\}$ be the maximum delay shift, and define $\bm{\mathrm{U}}^m_l = \sum_{k \in \mathcal{K}} h[l,k]e^{j2\pi \frac{k(m-l)}{M_DN_D}}\bm{\mathrm{C}}^k$. Then, the effective DD channel matrix $\bm{\mathrm{H}}$ can be written as
\begin{eqnarray}
	\label{Hmatrix}
	\begin{split}
		\bm{\mathrm{H}} =
		\scriptsize{
			\setlength{\arraycolsep}{3pt} 
			\begin{split}
				\begin{bmatrix}			  
					\bm{\mathrm{U}}^0_0&0&\cdots&0&\bm{\mathrm{U}}^0_{\mathfrak{L}}\bm{\mathrm{D}}&\cdots&\cdots&\bm{\mathrm{U}}^0_1\bm{\mathrm{D}}\\
					\vdots&\ddots&\ddots&\ddots&&\ddots&\ddots&\vdots\\
					\bm{\mathrm{U}}^{\mathfrak{L}-1}_{\mathfrak{L}-1}&\ddots&\ddots&\bm{\mathrm{U}}^{\mathfrak{L}-1}_{0}&\ddots&&\ddots&\bm{\mathrm{U}}^{\mathfrak{L}-1}_{\mathfrak{L}}\bm{\mathrm{D}}\\
					\bm{\mathrm{U}}^{\mathfrak{L}}_{\mathfrak{L}}&\ddots&\ddots&\ddots&\bm{\mathrm{U}}^{\mathfrak{L}}_{0}&\ddots&&0\\
					0&\ddots&\ddots&\ddots&\ddots&\ddots&\ddots&\vdots\\
					\vdots&\ddots&\ddots&\ddots&\ddots&\ddots&\ddots&0\\
					0&\cdots&0&\bm{\mathrm{U}}^{M_D-1}_{\mathfrak{L}}&\cdots&\cdots&&\bm{\mathrm{U}}^{M_D-1}_{0}\\
				\end{bmatrix}.
			\end{split}
		}	
	\end{split}
\end{eqnarray}

\begin{figure}[h]
	\centering
	\includegraphics[width = 220px]{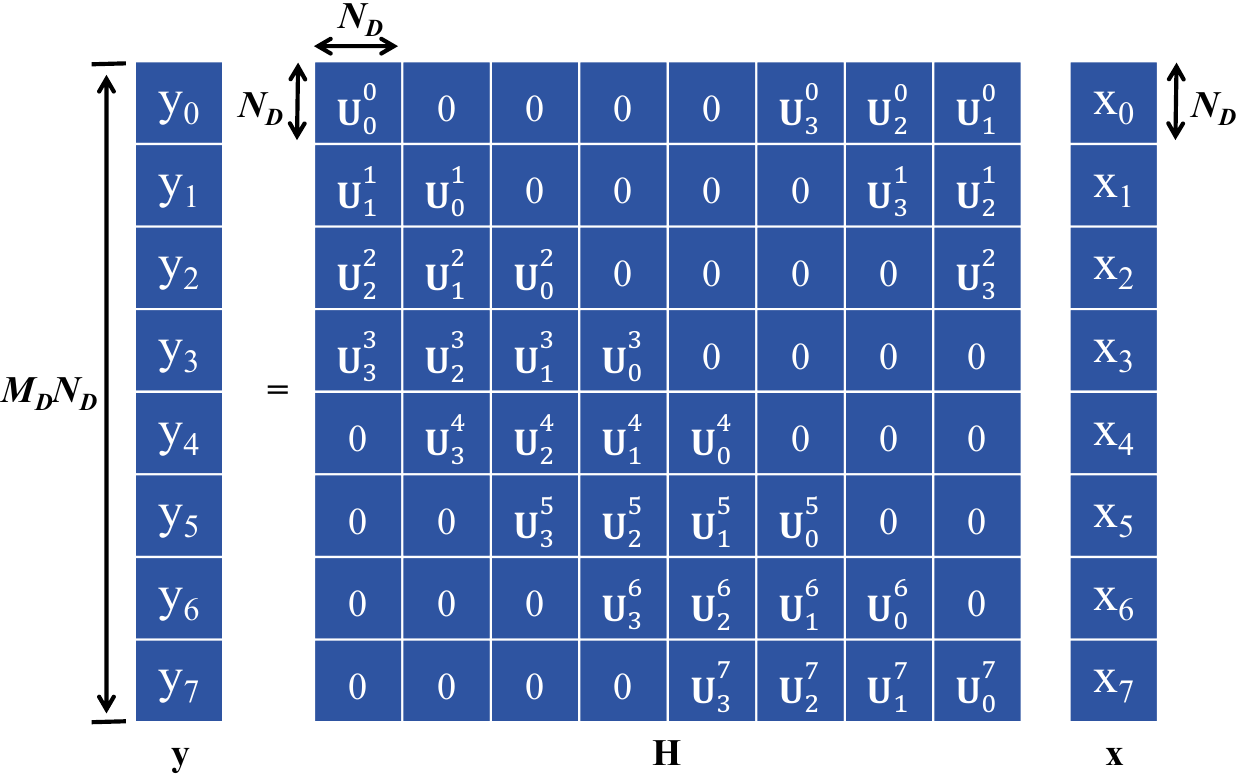}
	\caption{DD domain input-output relation when $M_D = N_D = 8$ and $\mathfrak{L}=3$.}
	\label{DDIO}
\end{figure}

For example, when $M_D=N_D=8$, and $\mathfrak{L}=3$, the DD domain input-output relation is shown in Fig. \ref{DDIO}.

In practical physical channels, achieving integer delay and Doppler shifts is challenging \cite{9738478}. Although this study focuses on integer delay and Doppler shifts, the power of off-grid interference can be effectively managed by adjusting the roll-off factor of the transmitted pulse. Specifically, increasing the roll-off factor of the SRRC pulse significantly mitigates the power of off-grid interference caused by fractional delays and Doppler shifts \cite{10562334}. However, this enhancement comes at the expense of a substantial increase in the bandwidth requirement.

\section{Channel estimation and data detection method for CDDM} 	\label{CE and DD}
\subsection{DD chirp based Channel Estimation Scheme} 	\label{CE}
In CDDM, the advantages of chirp signals and the DD channel are combined, allowing efficient exploitation of the sparse nature of the DD channel. This facilitates rapid estimation of delay and Doppler shifts, followed by accurate channel gain estimation through the correlation properties of DD chirps, achieving low-overhead and precise channel estimation.

\begin{figure*}[!h]
	\centering
	\begin{minipage}[b]{.95\linewidth}
		\centering
		\subfloat[][Using the $\bm{\mathrm \Phi}_{i_{\rho},DD}$ to generate DD domain signal pilot symbol.]{\includegraphics[width=.49\textwidth]{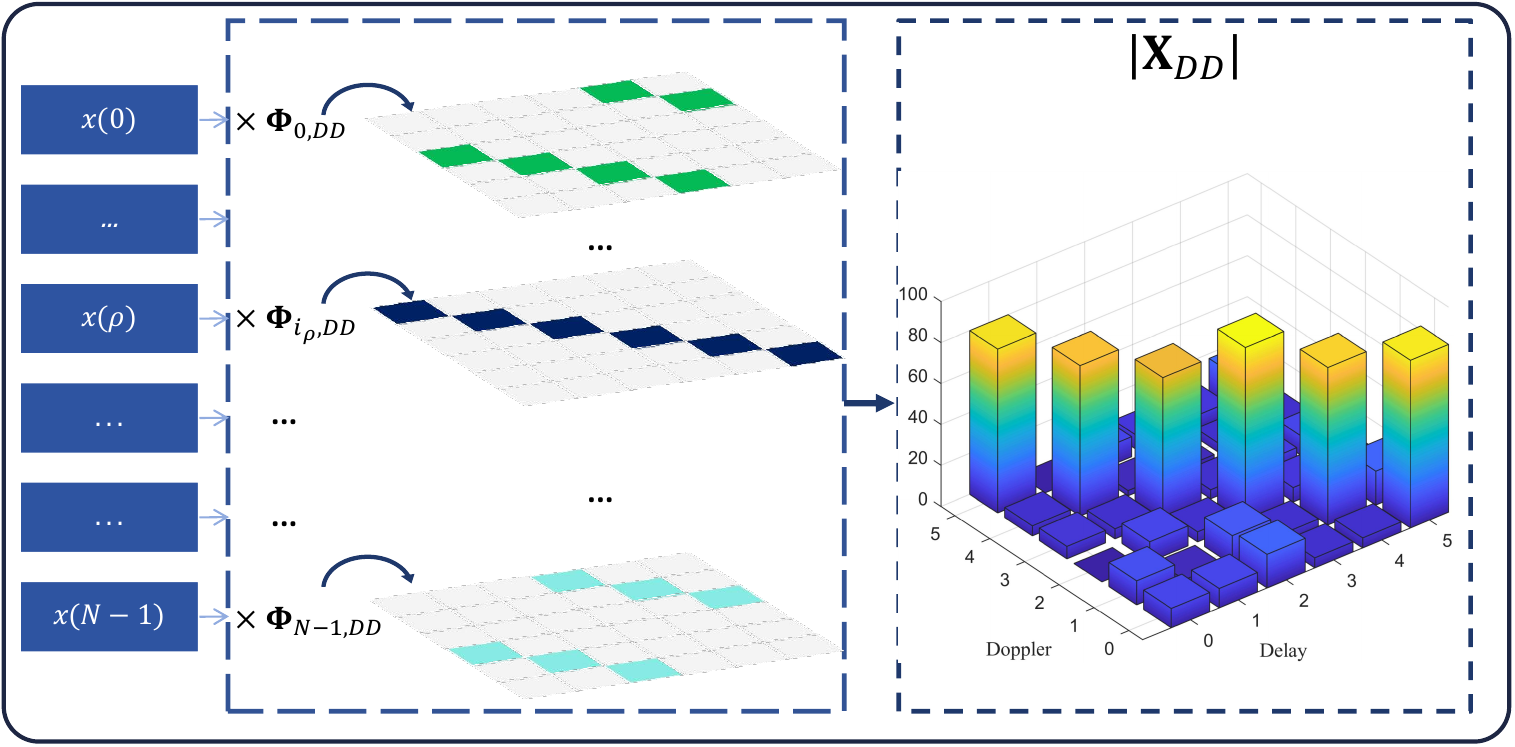}} \ 	
		\subfloat[][Modifying the $\bm{\mathrm \Phi}^{\lambda}_{i_{\rho},DD}$ to generate DD domain signal pilot symbol. Here ,this paper set $\lambda = 1$.]{\includegraphics[width=.49\textwidth]{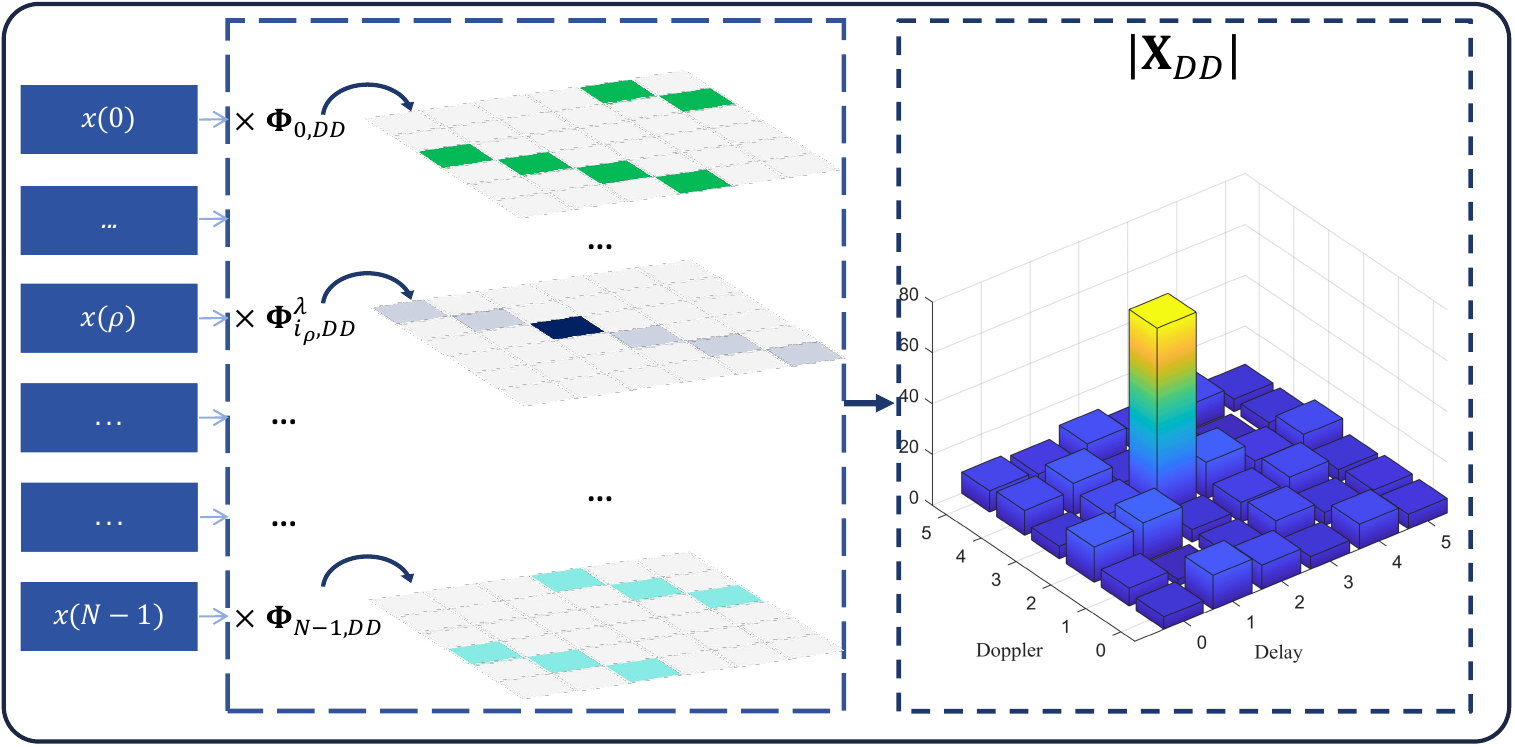}} \\
		\subfloat[][The distribution of pilots in the transmitted and received symbols.]{\includegraphics[width=.99\textwidth]{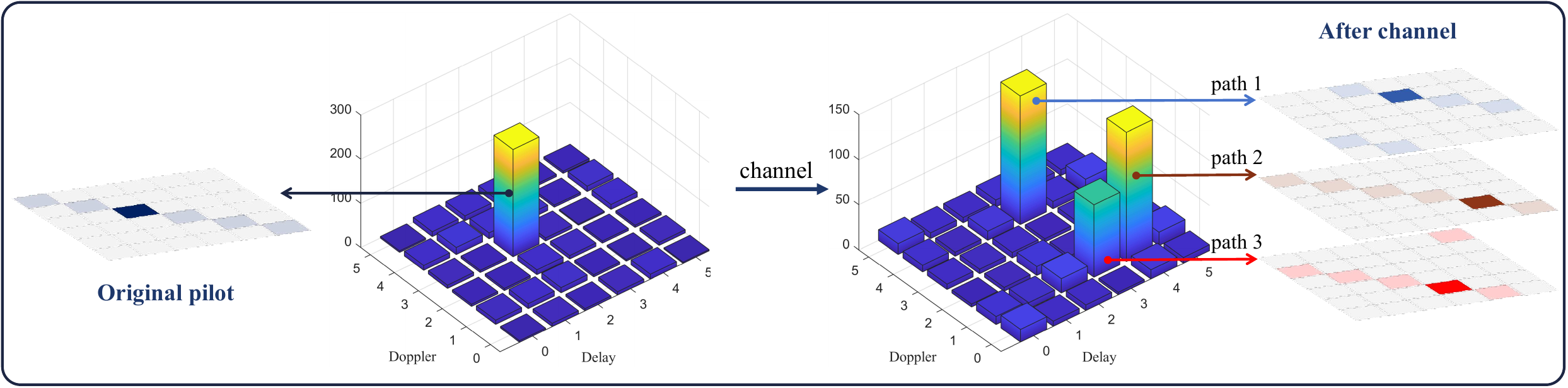}}
	\end{minipage}
	\caption{The pilot scheme of CDDM. Here, $\bm{\mathrm X}_{DD} =x(\rho)\bm{\mathrm \Phi}^{\lambda}_{i_{\rho},DD} + \sum_{i \neq i_{\rho}} x(i)\bm{\mathrm \Phi}_{i,DD}$.}
	\label{pilot}
\end{figure*}

The pilot scheme is illustrated in Fig. \ref{pilot}. Let us assume the $i_{\rho}$-th element of the $\bm x$ is designated as a pilot with value $x(\rho)$. After CZT, each CDDM frame can be viewed as a DD frame that contains diagonally placed delay-wise superimposed pilots. Although the continuous arrangement of multiple pilots offers better performance, it requires cyclic shifts of the pilot sequence for delay and Doppler shifts estimation during channel estimation, resulting in higher computational overhead compared to the threshold-based channel estimation with a single pilot. In Section \ref{OptimizedPilot}, an optimized threshold-based superimposed multi-pilot scheme based on CZT is proposed. For ease of understanding, this section uses a superimposed single pilot in this section to explain the channel estimation scheme. 

Let $i_{\rho}$-th CZT transformation matrix $\bm{\mathrm \Phi}_{i_{\rho},DD}$ retains  only one non-zero element, with all others set to zero. Here, this paper defines the number of non-zero elements in matrix $\bm{\mathrm \Phi}_{i_{\rho},DD}$ is $\lambda$. For example, when $M_D = N_D = 6$, the pilot is set at the position with index 2 in $\bm x$, which means $i_{\rho} = 2$. Then, the modified CZT matrix $\bm{\mathrm \Phi}^{\lambda}_{i_{\rho},DD}$ with $\lambda = 1$ is shown in Fig. \ref{pilot} (b), which is shown below: 
\begin{eqnarray}
	\begin{split}
		\label{pilotCZTmatrix}
		\bm{\mathrm \Phi}^{\lambda}_{i_{\rho},DD} = 
		\sqrt{6}
		\begin{bmatrix}			  
			0& 0& 0& 0& 0& 0\\
			0& 0& 0& 0& 0& 0\\
			0& 0& 0& \varphi_{i_{\rho}}(2)& 0& 0\\
			0& 0& 0& 0& 0& 0\\
			0& 0& 0& 0& 0& 0\\
			0& 0& 0& 0& 0& 0\\
		\end{bmatrix}.
	\end{split}
\end{eqnarray}

After passing through the DD channel, let us assume that there are 3 paths. The distribution of the pilot in the received signal $\bm{\mathrm Y}_{DD}$ is shown in Fig. \ref{pilot} (c). We define a threshold $\epsilon$, and when $\bm{\mathrm Y}[m,n] \geq \epsilon$, a pilot can be considered locates in this grid after a certain path. The delay and Doppler shift of this path can be quickly estimated by comparing the difference with the original pilot positions. At this point, the phase rotation introduced from this path can be obtained. The phase rotation of the $p$-th path in $m$-th multi-carrier symbol and the $n$-th sub-carrier is denoted as $\beta_p (m,n)$, which is given by:
\begin{eqnarray}
	\label{phaserotation}
	\small{
		\begin{split}
			\beta_p(m,n) = \begin{cases} e^{j2\pi\frac{k_p(m-l_p)}{M_DN_D}} &, \text{if} \  m \geq l_p\\ 
				e^{j2\pi\frac{k_p(m-l_p)}{M_DN_D}}e^{-j2\pi \frac{[n-k_p]_{N_D}}{N_D}} &, \text{if} \  m < l_p\end{cases}.
		\end{split}
	}
\end{eqnarray}

Subsequently, based on the correlation between DD chirps, the channel gain for the $p$-th path $h_p$ can be determined according to \eqref{FICZT}:
\begin{eqnarray}
	\label{esthp}
	\begin{split}
		h_p = \sum_{m=0}^{M_D-1} \frac{\bm{\mathrm Y}_{DD}^{\rho,p}(m) \varphi_{i_{\rho}}^*([m-l_p]_{M_D})}{x(\rho) \sqrt{N_D}\beta_{p,\rho}(m)}.
	\end{split}
\end{eqnarray}

Recall that the distribution of all data symbols in the DD domain have been saved, represented as $\Delta = \{ \Delta_0,\Delta_1,...,\Delta_{N-1}\}$. $\Delta_i = \{(m'_{i,0}, n'_{i,0}), (m'_{i,1}, n'_{i,1}),..., (m'_{i,M_D-1}, n'_{i,M_D-1})\}$ represent the distributions of all the non-zero entries associated to $x(i)$ after CZT. Therefore, $\bm{\mathrm Y}_{DD}^{\rho,p}$ represents $M_D$ elements, derived from $\Delta_{i_{\rho}}$, $l_p$ and $k_p$, used for correlation computation following the delay and Doppler shifts introduced by $p$-th path. The $m$-th element of $\bm{Y}_{DD}^{\rho,p}$ is given in \eqref{YDDp}. In a similar manner, $\beta_{p,\rho}$ denotes $M_D$ phase rotations derived from $\Delta_{i_{\rho}}$, $l_p$ and $k_p$. For the $i$-th element in $\bm x$, the $m$-th element of $\beta_{p,i}$ is given by
\begin{figure*}[ht]
	\centering 
	\hrulefill 
	\begin{eqnarray}		
		\begin{split}
			&\bm{\mathrm Y}_{DD}^{\rho,p}(m) =h_p\beta_{p,\rho}(m)x(\rho)\bm{\mathrm \Phi}^{\lambda}_{\rho,DD}[m'_{\rho,[m-l_p]_{M_D}}, n'_{\rho,[m-l_p]_{M_D}}]+ \sum_{p' \neq p}h_{p'}\beta_{{p'},\rho}(m)x(\rho)\bm {\mathrm \Phi}^{\lambda}_{\rho,DD}[m'_{\rho,[m-l_p]_{M_D}}+l_p-l_{p'},\\ 
			&n'_{\rho,[m-l_p]_{M_D}}+k_p-k_{p'}]+\sum_{p' \neq p} \sum_{i' \neq \rho}h_{p'}\beta_{p',i'}(m)x(i')\bm{\mathrm \Phi}_{i',DD}[m'_{\rho,[m-l_p]_{M_D}}+l_p-l_{p'}, n'_{\rho,[m-l_p]_{M_D}}+k_p-k_{p'}] + z(m).
		\end{split}
		\label{YDDp}
	\end{eqnarray}
	\hrulefill 
\end{figure*}
\begin{eqnarray}
	\label{betap_pilot}
	\begin{split}
		&\beta_{p,i}(m) =\\ 
		&\begin{cases} e^{j2\pi\frac{k_pm'_{i,[m-l_p]_{M_D}}}{M_DN_D}} &, \text{if} \  m \geq l_p \\ 
			e^{j2\pi\frac{k_pm'_{i,[m-l_p]_{M_D}}}{M_DN_D}}e^{-j2\pi \frac{n'_{i,[m-l_p]_{M_D}}}{N_D}} &, \text{if} \  m < l_p\end{cases}.
	\end{split}
\end{eqnarray}

\begin{figure}[h]
	\centering
	\includegraphics[width = 240px]{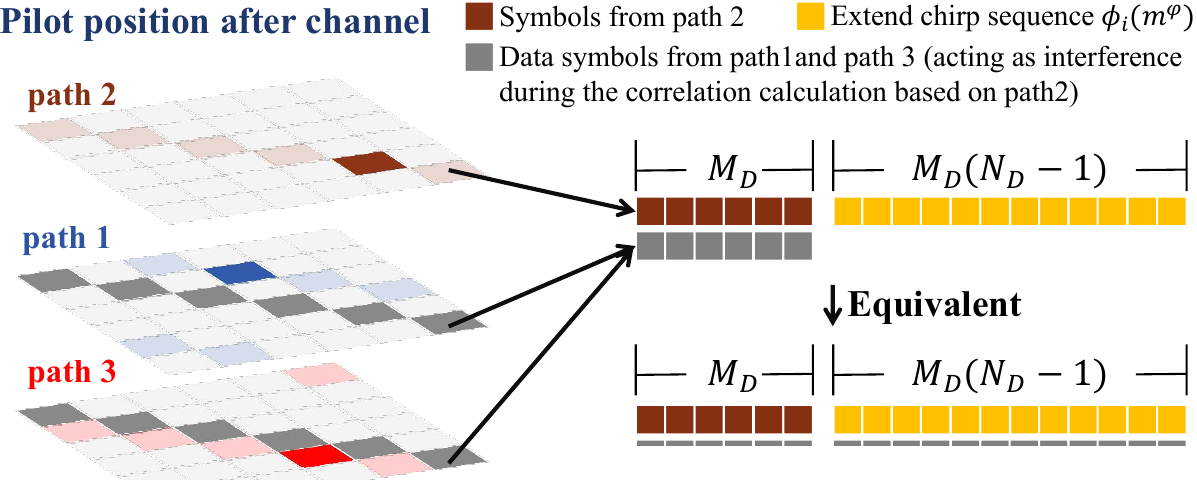}
	\caption{The strategy for extending the chirp sequence in correlation calculation to reduce the impact of interference.}
	\label{detectionmethod}
\end{figure}

Due to the doubly selective channel, non-orthogonal data symbols experience delay and Doppler shifts, causing interference during correlation calculations, as shown in Fig. \ref{detectionmethod}. For instance, when estimating path 2's channel gain, interference arises from paths 1 and 3 due to non-zero correlation. While large $x(\rho)$ values provide inherent anti-interference capability, additional mitigation remains necessary.


\begin{figure}[!h] 
	\centering
	\includegraphics[width = 200px]{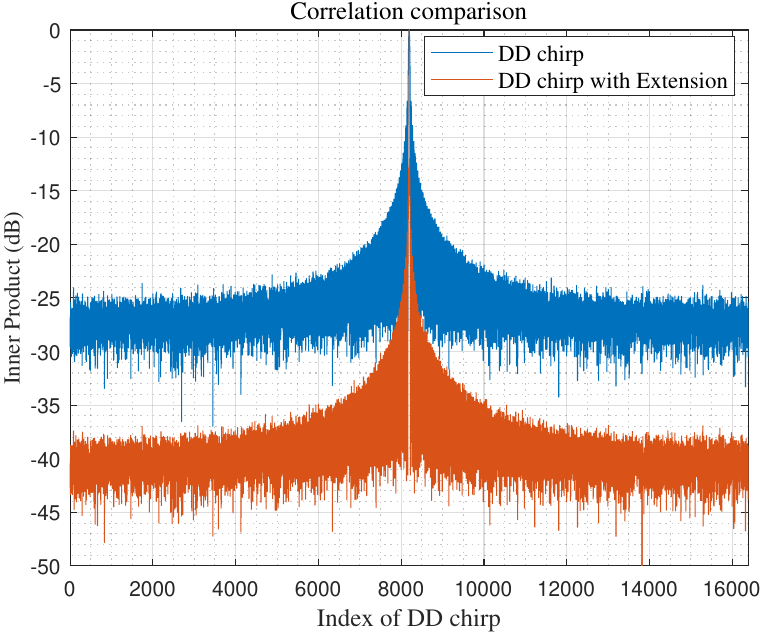}
	\caption{Correlation between $\frac{M_DN_D}{2}$-th data symbol and others in DD domain after spreading on DD chirps, with $M_D = 512, N_D = 32$.}
	\label{correlation}
\end{figure}

In \eqref{cztforxi}, the non-zero elements of each DD chirp are merely a segment of the complete chirp sequence, since the number of non-zero elements of each DD chirp is $M_D$, while the length of the full chirp sequence is $M_DN_D$. Given that all the DD chirp are known based on the precomputed, this paper considers completing the chirp at the receiver and performing correlation computation over a longer sequence, as shown in Fig. \ref{detectionmethod}. In Fig. \ref{detectionmethod}, the extended chirp sequence $\phi_{i}$ of length $M_D(N_D-1)$ can be expressed as
\begin{eqnarray}
	\label{extendchirp}
	\begin{split}
		\phi_{i}(m^{\varphi}) = &\sqrt{N_D} \varphi_{i}(m^{\varphi}), \\m^{\varphi}=M_D,M_D+&1,...,M_DN_D-1.
	\end{split}
\end{eqnarray}

The extended pilot chirp sequence is $\phi_{i_{\rho}}(m^{\varphi})$. With interference power evenly distributed across the $M_DN_D$-length sequence, longer sequences exhibit reduced interference fluctuations. After calculation, the extended sequence's contribution will be subtracted for more accurate channel estimation. The $p$-th path's channel gain is then estimated via \eqref{CEwithext}. Extending the DD chirp sequence reduces correlation between non-orthogonal chirps by 15 dB, as shown in Fig. \ref{correlation}. This lower correlation decreases interference from non-orthogonal symbols, enabling more accurate data recovery via correlation calculations.

\begin{figure*}[ht]
	\centering 
	\hrulefill 
	\begin{eqnarray}
		\label{CEwithext}
		\begin{split}
			h_p =\frac{\sum_{m=0}^{M_D-1} \frac{\bm{\mathrm Y}_{DD}^{\rho,p}(m)\varphi^*_{i_{\rho}}([m-l_p]_{M_D})}{x(\rho)\beta_{p,\rho}(m)} +\sum_{m^{\varphi}=M_D}^{M_DN_D-1}\phi_{i_{\rho}}(m^{\varphi}) \varphi^*_{i_{\rho}}(m^{\varphi}) }{\sqrt{N_D}(\lambda+(M_D(N_D-1)))} - \frac{M_D(N_D-1)}{\lambda+(M_D(N_D-1))}.
		\end{split}
	\end{eqnarray}
	\hrulefill 
	\begin{eqnarray}
		\setcounter{equation}{34}
		\label{YDDpi}
		\begin{split}
			&\bm{\mathrm Y}_{DD}^{i,p}(m) =h_p \beta_{p,i}(m) x(i)\bm{\mathrm \Phi}_{i,DD}[m'_{i,[m-l_p]_{M_D}}, n'_{i,[m-l_p]_{M_D}}] +\sum_{p' \neq p} \sum_{i' \neq i}h_{p'}\beta_{p',i'}(m)x(i')\bm{\mathrm \Phi}_{i',DD}[m'_{i,[m-l_p]_{M_D}} \\ 
			&+l_p-l_{p'}, n'_{i,[m-l_p]_{M_D}}+k_p-k_{p'}] + z(m).
		\end{split}
	\end{eqnarray}
	\hrulefill 
	\begin{eqnarray}
		\setcounter{equation}{35}
		\label{dataext} 
		\begin{split}
			x'(i) =\frac{1}{\sqrt{N_D}M_DN_D} \left( \sum_{m=0}^{M_D-1} \frac{\bm{\mathrm Y}_{DD}^{i,p_{max}}(m)\varphi^*_{i}([m-l_{p_{max}}]_{M_D})}{h_{p_{max}}\beta_{p_{max},i}(m)} +\sum_{m^{\varphi}=M_D}^{M_DN_D-1}\phi_{i}(m^{\varphi}) \varphi^*_i(m^{\varphi}) \right) - \frac{N_D-1}{N_D}.
		\end{split}
	\end{eqnarray}
	\hrulefill 
	
\end{figure*}


\begin{figure*}[ht]
	\centering
	
	\begin{minipage}[b]{.99\linewidth}
		\centering
		\subfloat[][$i$ = 0, $\alpha$ = 0]{\includegraphics[width=.27\linewidth]{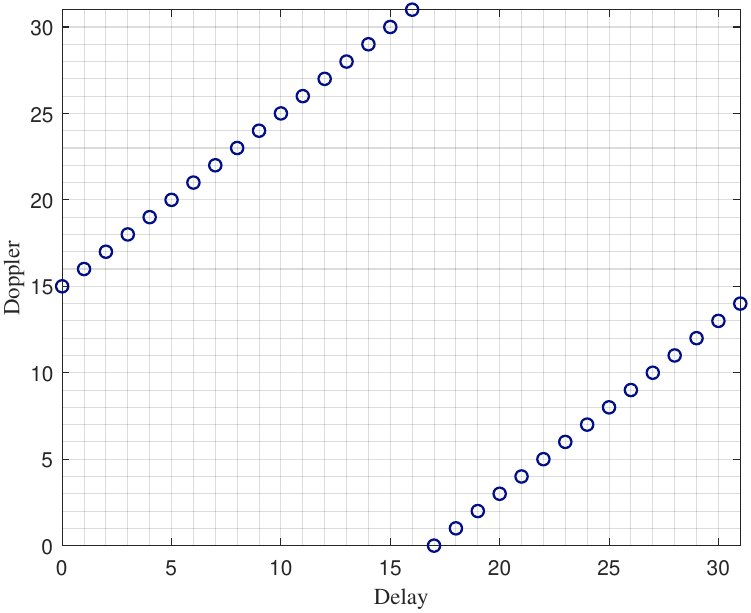}}
		\hspace{10pt}	
		\subfloat[][$i$ = 0, $\alpha$ = 3]{\includegraphics[width=.27\linewidth]{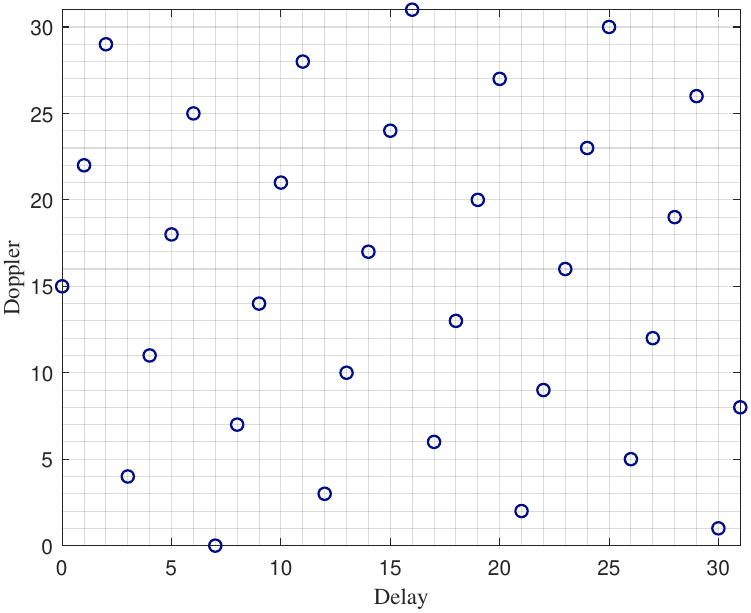}}
		\hspace{10pt}
		\subfloat[][$i$ = 0, $\alpha$ = 5]{\includegraphics[width=.27\linewidth]{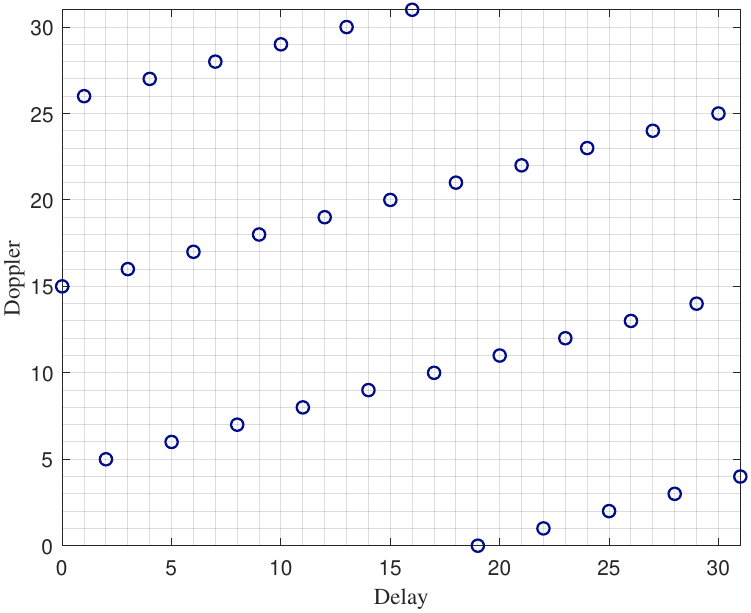}}\\
	\end{minipage}
	\caption{The distributions of non-zeros entries of $\mathcal{CZ}_{x(i)}^\xi$ when $M_D=N_D=32$.}
	\label{figCZTofDAFT}
\end{figure*}

\begin{figure}[ht]
	\centering
	\subfloat[][Distribution of pilots when $i$ = 0, $\alpha$ = 3, before passing through the channel.]{\includegraphics[width = 200px]{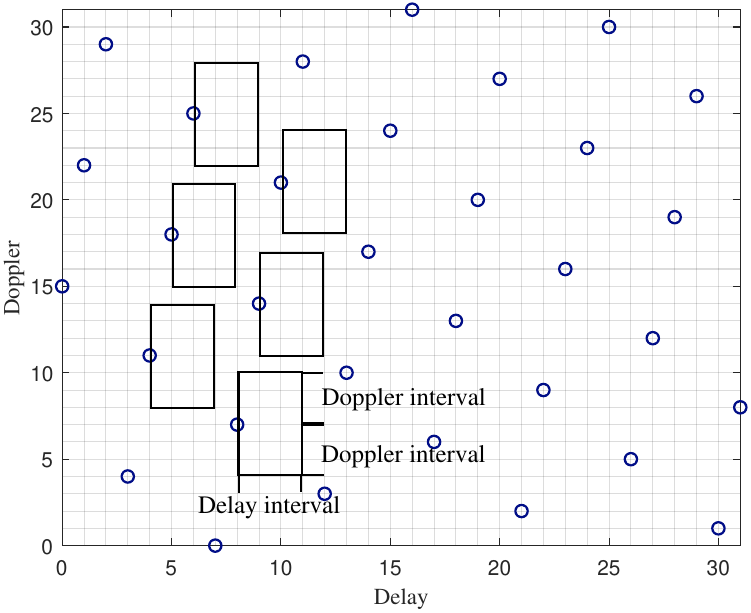}}	\\
	\subfloat[][Distribution of pilots when $i$ = 0, $\alpha$ = 3, after passing through the channel.]{\includegraphics[width = 200px]{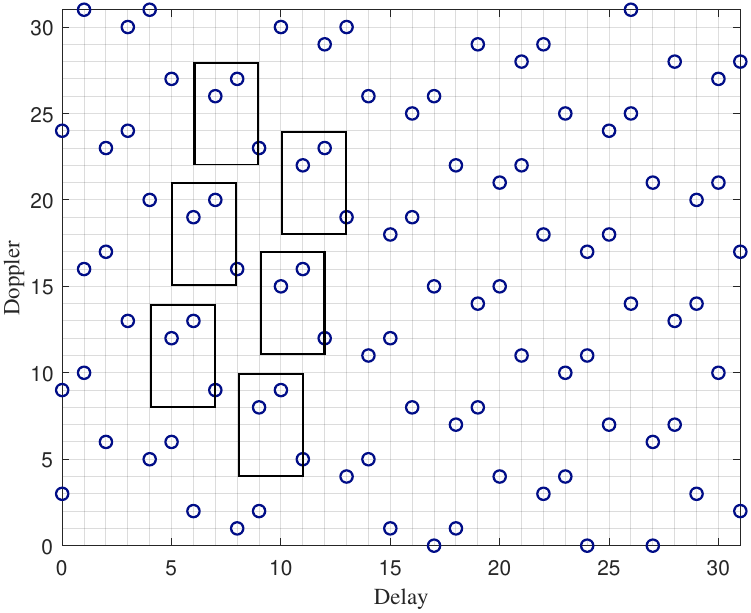}}
	\caption{The distribution of $\mathcal{CZ}^\xi_{x(\rho)}$ before and after passing through the channel when $M_D=N_D=32$ and $\lambda = M_D$. We employ 3-paths channel with delay shifts are 1, 2 and 3, while Doppler shifts are 1, 2, and -2.}
	\label{figCZTofDAFTpilot}
\end{figure}


\subsection{Data Detection Scheme} 	\label{DD}

As discussed, data symbols are recoverable via ICZT. However, doubly selective channels introduce interference during recovery. In high mobility scenarios (e.g., EVA channel \cite{3gpp.36.104}), path gains exhibit notable variations. The power delay profile (PDP) differences become pronounced after fading, creating distinct pulse variations in the DD channel matrix. Once the channel estimation is completed, data recovery can be achieved based on the DD domain channel information and the correlation between the DD chirps. Based on the DD domain channel matrix, the delay and Doppler shift of a certain path can be obtained, and determine the position of each data symbol under this path using $\Delta_i$.

Assume that the $i$-th data symbol, after passing through the $p$-th path, is distributed over the grids $\bm{\mathrm Y}_{DD}^{i,p}$ in the received DD signal, and $\bm{\mathrm Y}_{DD}^{i,p}(m)=\bm{\mathrm Y}_{DD}[m'_{i,m}+l_p, n'_{i,m}+k_p]$. At this point, the method for recovering the $i$-th data at the position corresponding to the $p$-th path based on the correlation of DD chirp is given by
\begin{eqnarray}
	\setcounter{equation}{33}
	\label{redatabeforeopt} 
	\begin{split}
		x'(i) = \sum_{m=0}^{M_D-1} \frac{\bm{\mathrm Y}_{DD}^{i,p}(m)\varphi^*_{i}([m-l_p]_{M_D})} {h_p\beta_{p,i}(m)M_D\sqrt{N_D}},
	\end{split}
\end{eqnarray}
where $x'(i)$ is the recovered $i$-th data symbol. $\bm{\mathrm Y}_{DD}^{i,p}$ is the grids where $i$-th data symbol located in after $p$-th path, which includes orthogonal data symbols from the $p$-th path, non-orthogonal data symbols from other paths and noise. We neglect the orthogonal components, the $\bm{\mathrm Y}_{DD}^{i,p}$ is expressed in \eqref{YDDpi}.

Recovering data using the path with maximum channel gain amplitude maximizes the denominator in \eqref{redatabeforeopt}, reducing interference impact. However, this approach offers limited improvement unless one path dominates significantly, which is a rare scenario even in high mobility cases. To better mitigate interference, CDDM employs extended chirp sequences for correlation calculations over longer sequences using \eqref{dataext}.

When transmit under equal paths, one needs to perform equalization before ICZT. For example, with the aid of channel estimation, algorithms such as least square (LS) can be used to reduce the impact of multipath. Recall that $\bm{\mathrm{x}}_{DD}$ is the vector representation of $\bm{\mathrm{X}}_{DD}$, and $\bm{\mathrm{y}}_{DD}$ is the vector representation of $\bm{\mathrm{Y}}_{DD}$, then, the recovered $\bm{\mathrm{x}}'_{DD}$ is
\begin{eqnarray}
	\setcounter{equation}{36}
	\label{LSXDD}
	\begin{split}
		\bm{\mathrm{x}}'_{DD} = \left( \bm{\mathrm{H}}^H \bm{\mathrm{H}} \right)^{-1}\bm{\mathrm{H}}^H \bm{\mathrm{y}_{DD}}.
	\end{split}
\end{eqnarray}

Subsequently, $\bm{\mathrm{x}}'_{DD}$ can be reshaped into a matrix to obtain $\bm{\mathrm{X}}'_{DD}$. Then, based on the correlation between DD chirps in the recovered $\bm{\mathrm{X}}'_{DD}$, accurate recovery can be achieved with ICZT. Evidently, this approach increases the complexity of the receiver, as it requires adding an additional ICZT step to the equalization algorithm.


\begin{figure}[ht]
	\centering
	\subfloat[][NMSE with different $\sigma_p$.]{\includegraphics[width = 200px]{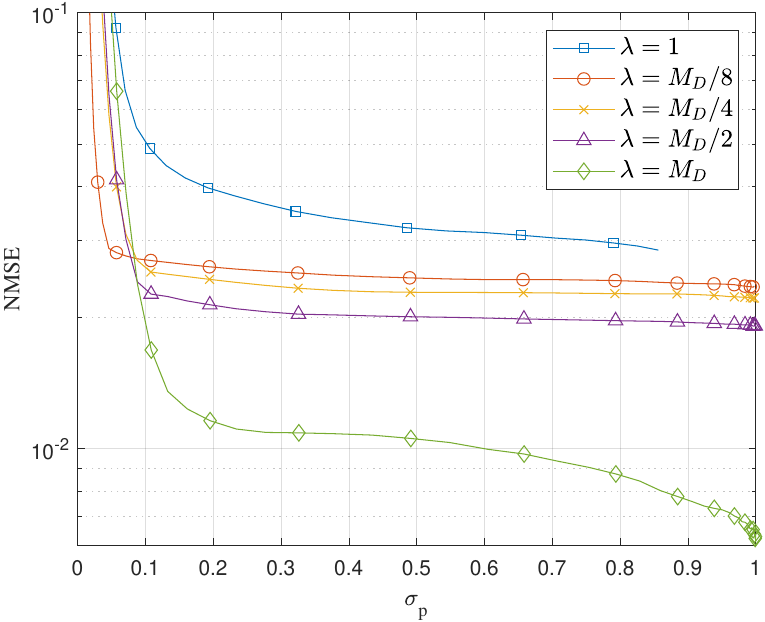}}	\\
	\subfloat[][BER with different $\sigma_p$.]{\includegraphics[width = 200px]{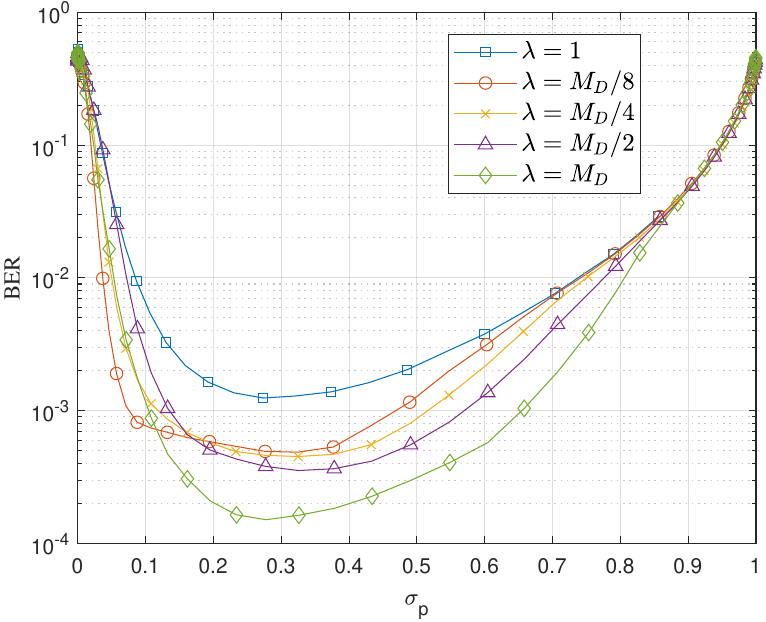}}
	\caption{NMSE and BER comparison with different $\sigma_p$ at $E_b/N_0$ of 12 dB. }
	\label{NMSEBERPAPR_AFDMCZT}
\end{figure}

\begin{figure*}[h]
	\centering
	\includegraphics[width = .95\textwidth]{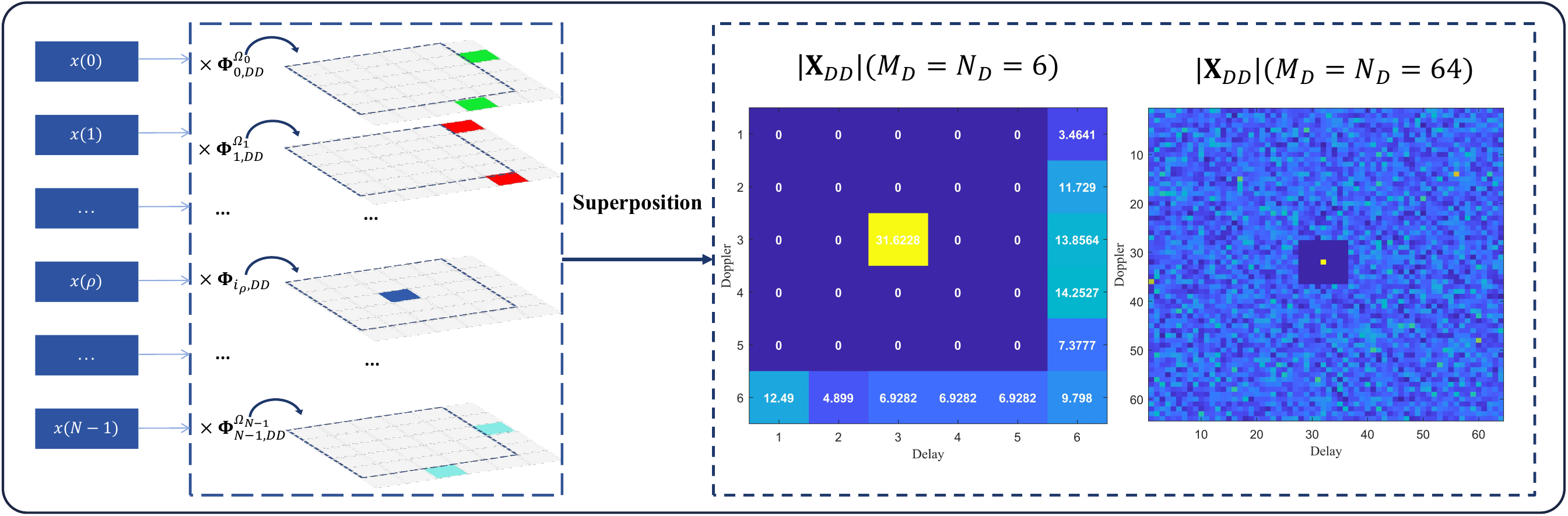}
	\caption{CDDM with EP.}
	\label{CDDMEP}
\end{figure*}

\section{Pilot scheme for CDDM based on the CZT} 	\label{OptimizedPilot}
In Section \ref{CE and DD}, the channel estimation techniques have been discussed, whereby using a superimposed single pilot was considered, as it simplifies the process of estimating delay and Doppler shifts. While a single pilot allows for simple and fast channel estimation, it may lead to poor performance. In conventional systems, superimposed pilots and embedded single pilot schemes represent two prevalent and effective pilot configurations.


This section presents CZT-based implementations of superimposed pilots (SP) and embedded single pilot (EP) for CDDM.

\subsection{CZT with chirps generated by DAFT}

The DAFT allows us to obtain orthogonal chirp sequences that contain parameters in their quadratic terms  \cite{9562168}. The $i$-th orthogonal chirp sequence of length $M_DN_D$ generated from DAFT is defined as
\begin{eqnarray}
	\begin{split}
		\xi_i(u) = e^{j2\pi (\mu_1u^2 + \frac{1}{M_DN_D}ui + \mu_2i^2)},\\
		u=0,1,...,M_DN_D-1,
	\end{split}
\end{eqnarray} 
where $\mu_1$ and $\mu_2$ are the DAFT parameters. Then, the CZT coefficient of $x(i)$ based on $\xi_i(u)$ can be written as
\begin{eqnarray}
	\label{cztforxibasedonAFDM}
	\begin{split} 
		\mathcal{CZ}_{x(i)}^\xi [m,n] = \frac{x(i)}{\sqrt{N_D}} \sum_{k=0}^{N_D-1} \xi_i (m+kM_D) e^{-j 2 \pi \frac{nk}{N_D}}.
	\end{split}
\end{eqnarray}

Here, parameters $\mu_1$ and $\mu_2$ are set as $\frac{2\alpha+1}{2M_DN_D}$ and $\frac{1}{2M_DN_D}$, respectively, where $\alpha$ is the maximum integer Doppler shift. Then, \eqref{cztforxibasedonAFDM} can be further simplified to \eqref{cztforxibasedonAFDMsimplified}.

In \eqref{cztforxibasedonAFDMsimplified}, $f^\xi_i=(2\alpha+1)m+\frac{2\alpha+1}{2}M_D+i-n$, similar with $\mathcal{CZ}_{x(i)}$, $\mathcal{CZ}_{x(i)}^\xi$ is also a sparse matrix shown in Fig. \ref{figCZTofDAFT}. We can also achieve CZT quickly through pre-computation, the fast implementation with $\xi_i(u)$ can be defined as
\begin{eqnarray}
	\setcounter{equation}{40}
	\label{FASTCZTAFDM}
	\begin{split}
		\mathcal{CZ}^\xi_{x(i)} &= x(i)\bm{\mathrm \Xi}_{i,DD}, 
	\end{split}
\end{eqnarray}
where $\bm{\mathrm \Xi}_{i,DD} = \frac{1}{\sqrt{N_D}}\bm{\mathrm \Xi}_{i} \bm{\mathrm F}_{N_D}$, $\bm{\mathrm \Xi}_{i}$ is defined as
\begin{eqnarray}
	\begin{split}
		\bm{\mathrm \Xi}_i = 
		\small{
			\setlength{\arraycolsep}{3pt} 
			\begin{bmatrix}		
				\xi_i (0) & \xi_i (M_D) &  ... & \xi_i ((N_D-1)M_D)  \\
				\xi_i (1) & \xi_i (M_D + 1) & ... & ...\\
				...& ... &...& ...\\
				\xi _ i (M_D-1) & \xi _ i (2M_D-1)  & ... & \xi _ i (M_DN_D -1)
			\end{bmatrix}
		}.
	\end{split}
\end{eqnarray}

From \eqref{cztforxibasedonAFDMsimplified}, it is evident that the distribution of non-zero entries of $\mathcal{CZ}^\xi_{x(i)}$ also exhibits cyclic shifts along the Doppler axis, resulting in overlap. Likewise, the overlapping regions still maintain orthogonality. Therefore,the distribution position $\Delta_i$ of each DD chirp can also be stored and efficiently implement the ICZT.

\subsection{Improved Superimposed Delay-Wise Multiple Pilots for CDDM based on CZT}

As shown in Fig. \ref{figCZTofDAFT}, when CZT is applied based on orthogonal chirp from DAFT, the non-zero elements of each DD chirp are more sparsely distributed. In this distribution, within a limitation of delay offset, the non-zero elements are sufficiently spaced to prevent interference between them, which can be considered as a certain form of guard interval between pilots. 

When the parameters of the orthogonal chirp generated by DAFT are set based on the maximum Doppler shift, if $M_D = N_D$, the $\bm{\mathrm \Xi}_{\rho,DD}$ inherently has a Doppler guard interval. Therefore, by setting pilots in this way, the pilots do not interfere with each other within a specific delay range, allowing for direct estimation of delay and Doppler shift through the threshold method. For example, in Fig \ref{figCZTofDAFTpilot}, the black boxes represent the guard interval between pilots. 

Since the delay interval in this strategy is not based on the maximum delay, pilot symbols may still interfere under high delay. To mitigate this, the number of non-zero elements $\lambda$ in matrix $\bm{\mathrm \Xi}_{i_{\rho},DD}$ can be reduced to increase the delay interval. 

Based on the orthogonal chirp generated from DAFT, the transmitted symbols in the DD domain with pilots for CDDM can be expressed as $\bm{\mathrm X}^{\xi}_{DD}$, which can be written as
\begin{eqnarray}
	\setcounter{equation}{42}	
	\label{XDDAFDM}
	\begin{split}
		\bm{\mathrm X}^{\xi}_{DD} =x(\rho)\bm{\mathrm \Xi}^{\lambda}_{i_{\rho},DD} + \sum_{\ \ i \neq i_{\rho}} x(i)\bm{\mathrm \Xi}_{i,DD}.
	\end{split}
\end{eqnarray}

\begin{figure*}[!h]
	\centering 
	\hrulefill 
	\vspace*{2pt} 
	\begin{eqnarray}
		\setcounter{equation}{39}
		\label{cztforxibasedonAFDMsimplified}			
		\begin{split} 
			\mathcal{CZ}_{x(i)}^\xi [m,n] &= \frac{x(i)}{\sqrt{N_D}} \sum_{k=0}^{N_D -1} e^{j2\pi (\frac{2\alpha+1}{2M_DN_D}(m+k M_D)^2 + \frac{1}{M_DN_D}(m+k M_D)i + \frac{1}{2M_DN_D}i^2)} e^{-j 2 \pi \frac{nk}{N_D}}\\
			&= \frac{x(i)}{\sqrt{N_D}} e^{j2\pi (\frac{2\alpha+1}{2M_DN_D}m^2+\frac{mi}{M_DN_D}+\frac{i^2}{2M_DN_D})}\sum_{k=0}^{N_D-1}e^{j2\pi\frac{kf^\xi_i}{N_D}} =\begin{cases} \sqrt{N_D}x(i)\xi_i(m) &,{\text{if} \  [ f^\xi_i ]_{N_D} = 0}\\ 0 &, \text{otherwise}\end{cases}.
		\end{split}
	\end{eqnarray}
	\hrulefill 
	\begin{eqnarray}
		\setcounter{equation}{43}
		\label{CEwithextAFDM} 
		\begin{split}
			h_p =\frac{\sum_{m=0}^{M_D-1} \frac{\bm{\mathrm Y}_{DD}^{\rho,p}(m)\xi^*_{i_{\rho}}([m-l_p]_{M_D})}{x(\rho) \beta_{p,\rho}(m)} +\sum_{m^{\xi}=M_D}^{M_DN_D-1}\phi^\xi_{i_{\rho}}(m^{\xi}) \xi^*_{i_{\rho}}(m^{\xi})}{\sqrt{N_D}(\lambda+(M_D(N_D-1)))} - \frac{M_D(N_D-1)}{\lambda+(M_D(N_D-1))}.
		\end{split}
	\end{eqnarray}
	\hrulefill 
	\begin{eqnarray}
		\setcounter{equation}{47}
		\label{datadetectionEP} 
		\begin{split}
			x'(i) =\frac{\sum_{m=0}^{M_D-1} \left(\bm{\mathrm X}'_{DD}[m'_{i,m},n'_{i,m}]\varphi^*_{i}([m-l_{p_{max}}]_{M_D} \right) +\sum_{m^{\varphi}=M_D}^{M_DN_D-1}\phi_{i}(m^{\varphi}) \varphi^*_i(m^{\varphi}) }{\sqrt{N_D}(M_DN_D-\Omega_i)} - \frac{M_D(N_D-1)}{M_DN_D-\Omega_i}.
		\end{split}
	\end{eqnarray}
	\hrulefill 
\end{figure*}

Subsequently, the distribution positions of each data and pilots can be stored in $\Delta$, followed by DD domain multi-carrier modulation. At the receiver, channel estimation and data detection are performed using the correlation of DD chirps. 

The formula of channel estimation is shown in \eqref{CEwithextAFDM}, where $\phi^\xi_i$ is the extend chirp sequence of $\xi_i$, i.e., 
\begin{eqnarray}
	\setcounter{equation}{44}
	\label{extendchirpafdm}
	\begin{split}
		\phi^\xi_{i}(m^{\xi}) = &\sqrt{N_D} \xi_{i}(m^{\xi}), \\m^{\xi}=M_D,M_D+&1,...,M_DN_D-1.
	\end{split}
\end{eqnarray}

In this case,  delay and Doppler shifts can be rapidly estimated using threshold within such a guard interval. However, this approach has limitations, as shown in Fig. \ref{figCZTofDAFTpilot}, the fixed pilot spacing, which is not based on maximum delay shift, permits inter-pilot interference at very high delays. While setting some of the non-zero elements in $\bm{\Xi}_{i_{\rho},DD}$ can increase the delay guard interval, this compromises both NMSE and BER performance.

\vspace{-0.7em}	
\subsection{Superimposed pilot Power Design}
There is no doubt that the low power of pilot symbols can lead to a degradation in channel estimation performance, which subsequently results in a deterioration of BER performance. As mentioned in \cite{9539066}, for superimposed pilot schemes, increasing pilot power is not always beneficial. While higher pilot power improves channel estimation performance, it can paradoxically degrade BER performance.

In CDDM, each grid value is a superposition of multiple complex numbers, which may lead to random amplification or cancellation. Additionally, pilot symbols, being complex values after CZT, are also subject to amplification or cancellation, as shown in Fig. \ref{pilot} (a). Therefore, the definition for pilot power proportion in this paper differs from that in \cite{9539066}. In CDDM, the proportion of pilot power in the total power is defined as $\sigma_d$, and the proportion of data power in the total power is defined as $\sigma_p$, which are given by
\begin{eqnarray}
	\label{powerdataXDD}
	\sigma_d =\frac{\sum_{m=0}^{M_D-1} \sum_{n=0}^{N_D-1}|\sum_{i \neq i_{\rho}} x(i)\bm{\mathrm \Xi}_{i,DD}[m,n]|^2}{\sum_{m=0}^{M_D-1} \sum_{n=0}^{N_D-1}|\bm{\mathrm X}^{\xi}_{DD}[m,n]|^2},\\
	\label{powerpilotXDD}
	\sigma_p =\frac{\sum_{m=0}^{M_D-1} \sum_{n=0}^{N_D-1}|x(\rho)\bm{\mathrm \Xi}^{\lambda}_{i_{\rho},DD}[m,n]|^2}{\sum_{m=0}^{M_D-1} \sum_{n=0}^{N_D-1}|\bm{\mathrm X}^{\xi}_{DD}[m,n]|^2}.
\end{eqnarray}

In Fig. \ref{NMSEBERPAPR_AFDMCZT}, when $E_b/N_0$ is equal to 12 dB, the NMSE and BER performances under different values of $\sigma_p$ and $\lambda$ are compared. Fig. \ref{NMSEBERPAPR_AFDMCZT} (a) compares the NMSE performance under different values of $\lambda$. As expected, with the increase of $\sigma_p$, the performance of NMSE improves progressively. For the same $\sigma_p$, increasing the number of pilot symbols results in better NMSE performance due to the stronger anti-interference capability of longer pilot sequences. As seen from  Fig. \ref{NMSEBERPAPR_AFDMCZT} (b), although a larger value of $\sigma_p$ improves NMSE performance, excessively large $\sigma_p$ actually degrades the BER performance. Based on the results in Fig. \ref{NMSEBERPAPR_AFDMCZT} (b), the optimal BER performance is achieved when $\sigma_p$ is approximately 0.3 when the ICZT is used for data recovery. This aligns with the findings in \cite{9539066}, which suggest that the performance is optimal when the power of superimposed pilots account for 30\% of the total power.

\subsection{Embedded pilot scheme for CDDM based on CZT} 	\label{Optimized-e-Pilot}

Embedded pilots (EP) generally demonstrate superior performance compared to superimposed pilots \cite{8671740} with spectral efficiency loss. While CDDM superimposes DD domain matrices for data generation, the precomputable CZT matrices permit guard interval insertion for EP through targeted modifications. We first select the pilot position (typically at the center of the DD matrix), then calculate the guard interval range based on the maximum velocity and maximum delay. In the CZT matrices $\bm{\mathrm \Phi}_{i,DD}$ for $i=0,1,...,N-1$, all the elements in the guard interval are set to zero. After superimposing, the $|\bm{\mathrm X}_{DD}|$ with EP can be get, as shown in Fig. \ref{CDDMEP}.

Through precomputation, the nullified non-zero elements $\Omega_i$ in $\bm{\mathrm \Phi}_{i,DD}$ can be counted, yielding the modified DD chirp matrix $\bm{\mathrm \Phi}^{\Omega_i}_{i,DD}$ in Fig. \ref{CDDMEP}.

This scheme enables efficient threshold-based channel estimation \cite{8671740}, though it alters DD chirp correlations due to the change of non-zero elements. Therefore, while EP improves channel estimation accuracy, subsequent equalization before ICZT remains essential for reliable data detection. Given the equalized DD domain received symbol matrix $\bm{\mathrm{X}}'_{DD}$, the
$i$-th recovered data symbol is expressed in \eqref{datadetectionEP}.

\section{Performance Evaluation and Discussions} 	\label{Simulation}
In this section, the proposed CDDM is compared with several classical and emerging waveforms, including OFDM, OCDM, CCDT, AFDM, OTFS, and ODDM. 

In Table I, the SFFT stands for Symplectic Finite Fourier Transform. Further, $b$ represents the number of iterations, $S$ denotes the number of non-zero elements per row in the channel matrix, and $\mathbb{A}$ refers to the modulation alphabet. From this comparison, it can be observed that the CDDM exhibits a relatively high transmitter complexity. Although the complexity expression appears tractable, practical implementations involve additional overheads. Specifically, DD multicarrier modulation requires upsampling before domain conversion to preserve symbol interval \cite{9829188}, leading to non-negligible cost. Moreover, pulse shaping with g(t) introduces extra computational burden. Finally, the fast implementation of CZT relies on pre-computation. While the CSR strategy substantially reduces the complexity of CZT, extra computations remain and hardware memory is needed to store pre-computed data.

\begin{figure}[ht]
	\centering
	\subfloat[][BER in EVA channel model.]{\includegraphics[width = 200px]{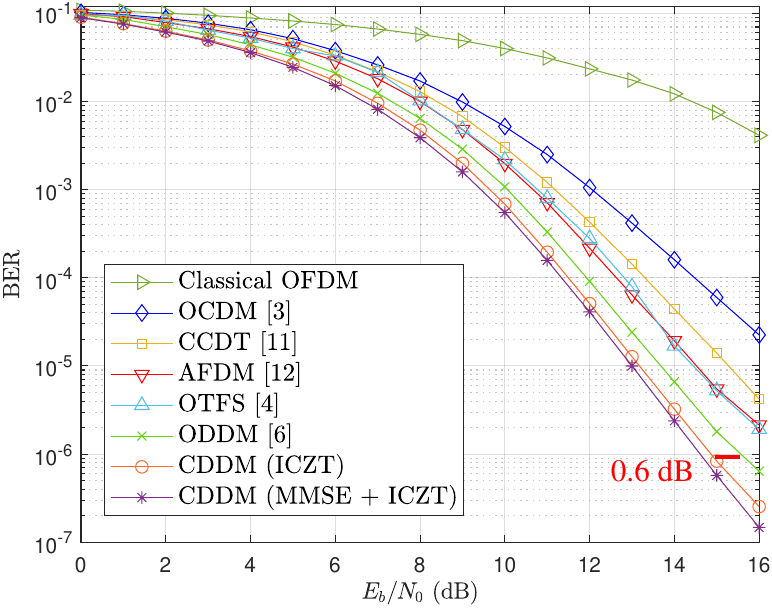}}	\\
	\subfloat[][BER in channel with uniform PDP.]{\includegraphics[width = 200px]{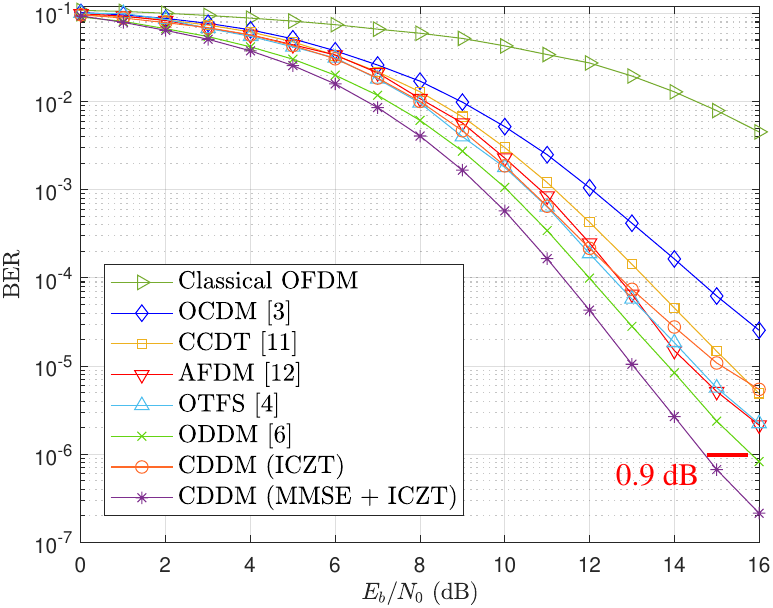}}
	\caption{BER comparison with perfect channel estimation at 500 km/h. }
	\label{CMPBERPerfectCSI}
\end{figure}

\begin{figure}[ht]
	\centering
	\subfloat[][BER in EVA channel model.]{\includegraphics[width = 200px]{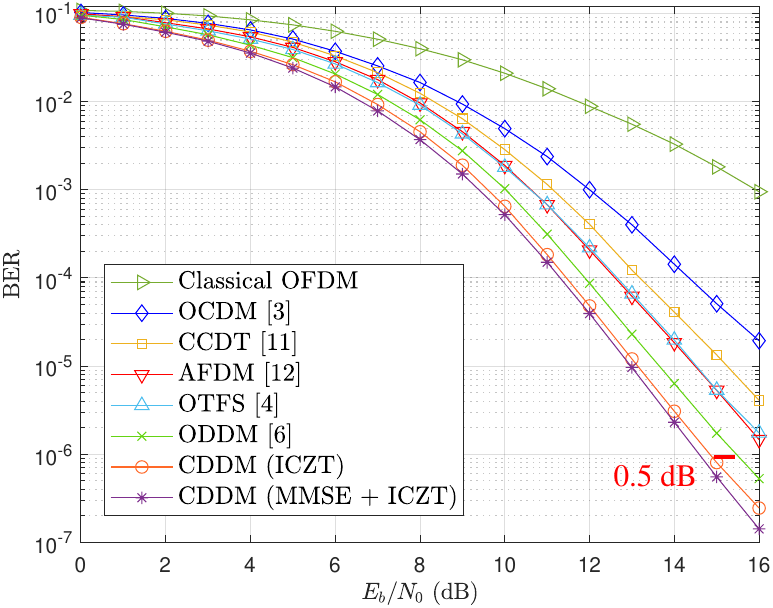}}	\\
	\subfloat[][BER in channel with uniform PDP.]{\includegraphics[width = 200px]{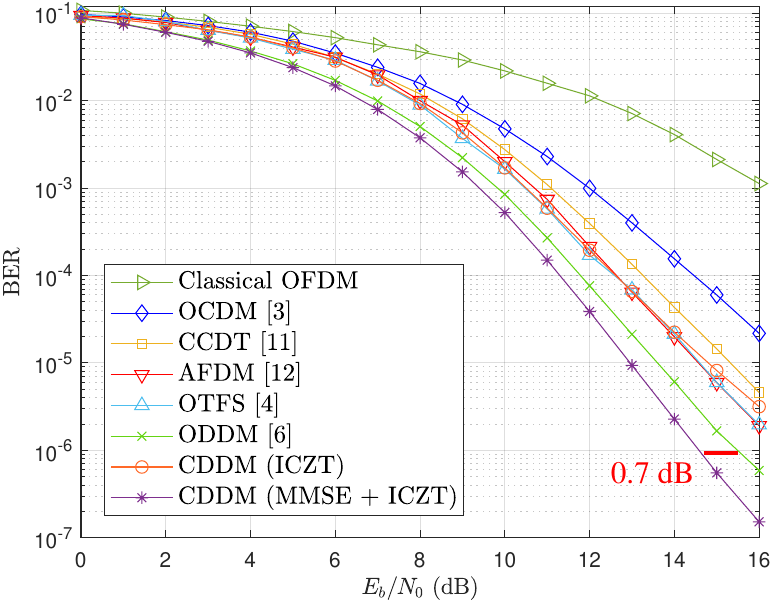}}
	\caption{BER comparison with perfect channel estimation at 120 km/h. }
	\label{CMPBERPerfectCSI120}
\end{figure}

\begin{table*}[!t]
	\label{cmpwaaveform}
	\centering
	\begin{threeparttable}	
		\normalsize
		\centering
		\caption{Comparison with other various modulation waveforms} 
		\label{complexity}
		\renewcommand{\arraystretch}{1} 
		\footnotesize
		\begin{tabular}{|c|c|c|c|c|} 		
			\hline 
			Modulation &
			\begin{tabular}{c}Transform Domain\end{tabular} &
			\begin{tabular}{c}Core Transformation\end{tabular} &
			\begin{tabular}{c}Modulation Complexity\end{tabular} &
			\begin{tabular}{c}Detector and Complexity\end{tabular}\\
			\hline
			OFDM& TF & DFT & $\mathcal{O}(N\log N)$ & \begin{tabular}{c} MMSE, $\mathcal{O}(N^3)$ \end{tabular} \\
			
			\hline
			OCDM \cite{7523229}& TF & DFnT & $\mathcal{O}(N\log N+2N)$ & \begin{tabular}{c} MMSE, $\mathcal{O}(N^3)$ \end{tabular}  \\
			\hline
			CCDT \cite{9284570}& TF & DFT & $\mathcal{O}(2N\log N+N)$ & \begin{tabular}{c} MMSE, $\mathcal{O}(N^3)$ \end{tabular}  \\
			\hline
			AFDM \cite{9562168}& TF & DAFT & $\mathcal{O}(N\log N+2N)$ & \begin{tabular}{c} MMSE, $\mathcal{O}(N^3)$ \end{tabular}  \\
			\hline
			OTFS \cite{7925924}& DD &\begin{tabular}{c} DZT or SFFT \end{tabular} & $\mathcal{O}(N\log N_D)$ or $\mathcal{O}(N\log N)$ & \begin{tabular}{c} MP, $\mathcal{O}(b N S |\mathbb{A}|)$ \end{tabular} \\
			\hline
			ODDM \cite{9829188}& DD & DFT & $\mathcal{O}(N\log N_D)$ & \begin{tabular}{c} MP, $\mathcal{O}(b N S |\mathbb{A}|)$ \end{tabular} \\
			\hline
			\begin{tabular}{c}CDDM(Proposed)\end{tabular} & DD & \begin{tabular}{c}CZT and DFT \end{tabular}  & $\mathcal{O}(2M_DN+N\log N_D)$ & \begin{tabular}{c} ICZT, $\mathcal{O}(2M_DN)$ \end{tabular} \\
			\hline
		\end{tabular}
		
		\begin{tablenotes}
			\footnotesize
			\item *In this table, $b$ represents the number of iterations, $S$ denotes the number of non-zero elements per row in the channel matrix, and $\mathbb{A}$ is the modulation alphabet.
		\end{tablenotes}
	\end{threeparttable}
	
\end{table*}

We also compare the NMSE and BER performance of CDDM with OFDM, OCDM, CCDT, AFDM, OTFS, and ODDM. Simulation parameters are detailed in Table II. Meanwhile, for the ODDM modulation, a SRRC pulse with a roll-off factor of 0.1 is employed as $a(t)$, therefore, the bandwidth of CDDM is about 8.45M Hz. We select four paths from the EVA channel for simulation, with delays of 0, 310, 710, and 1090 ns, and corresponding PDP values of 0, –3.6, –9.1, and –7.0 dB, respectively. 

\begin{table}[h]
	\centering
	\normalsize
	\caption{Simulation Parameters} 
	\label{samples}
	\vspace{5pt}
	\begin{tabular}{cc}
		\toprule[1pt] 
		Parameter &Value\\
		\midrule[1pt]
		Number of CDDM symbols $M_D$   &512\\
		Number of subcarriers $N_D$   &32\\
		Symbol duration $T$ & 66.67 $\mu$s\\
		Carrier frequency   & 5 GHz \\
		Moving Speed & 500 km/h and 120 km/h\\
		Modulation scheme & QPSK \\
		\bottomrule[1pt]
	\end{tabular}
\end{table}


In Fig. \ref{CMPBERPerfectCSI} (a), waveforms operating in the DD domain exhibit an advantage in EVA model, thanks to the high diversity gain. For data recovery, MMSE detector is adopted for OFDM, OCDM, AFDM, CCDT, OTFS and ODDM. For AFDM, its excellent path independence in channel matrix allows for effective mitigation of off-grid interference, thus achieving comparable performance to OTFS. Both ODDM and CDDM perform multi-carrier modulation in the DD domain and utilize SRRC pulse for pulse shaping, leading to improved signal-channel coupling and better BER performance. As seen in the figure, at the BER of $10^{-6}$, CDDM achieves approximately 1 dB performance gain compared to ODDM with only ICZT. This improvement is attributed to the DD chirps, allowing different data symbols to be well separated based on the correlation. Pre-ICZT MMSE equalization enhances BER performance through interference cancellation to improved calculation accuracy. In equal channel, standalone ICZT fails to achieve satisfactory data recovery due to correlation interference, as shown in Fig. \ref{CMPBERPerfectCSI} (b). Equalization prior to ICZT becomes essential for attaining good performance in such scenarios. Fig. \ref{CMPBERPerfectCSI120} compares the BER performance at 120 km/h. In the EVA model, although the performance of other waveforms improves, CDDM still achieves about 0.5 dB gain. Moreover, under the channel with uniform PDP, CDDM maintains approximately a 1 dB improvement with additional MMSE assistance.

\begin{figure}[ht]
	\centering
	\includegraphics[width = 200px]{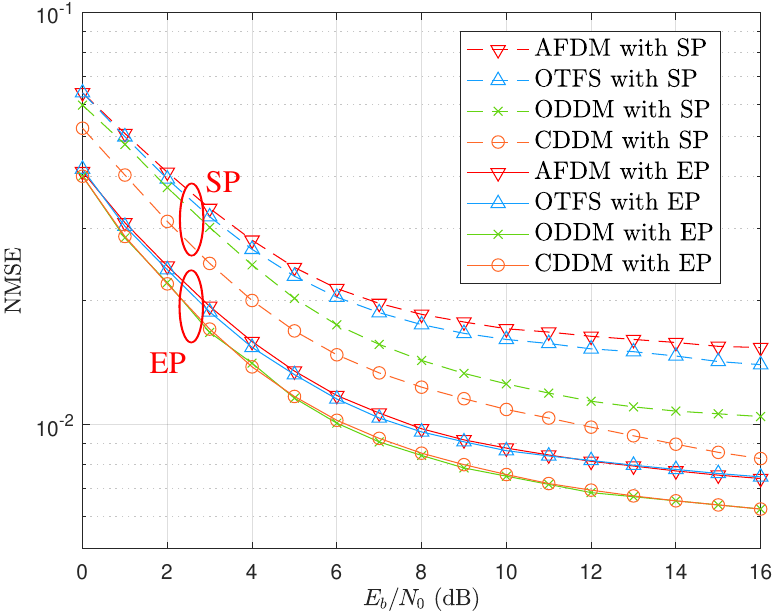}
	\caption{NMSE comparison with different DD pilot schemes. }
	\label{NMSE}
\end{figure}

\begin{figure}[ht]
	\centering
	\subfloat[][]{\includegraphics[width = 200px]{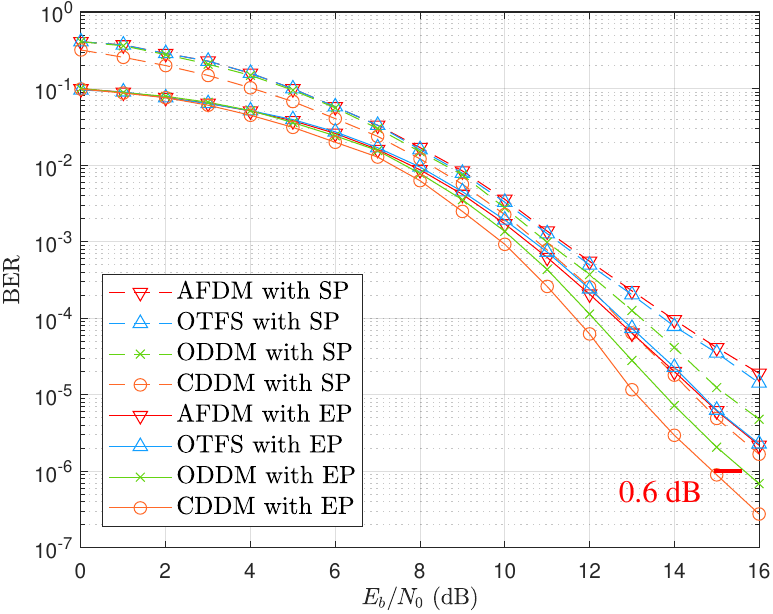}}	\\
	\subfloat[][]{\includegraphics[width = 200px]{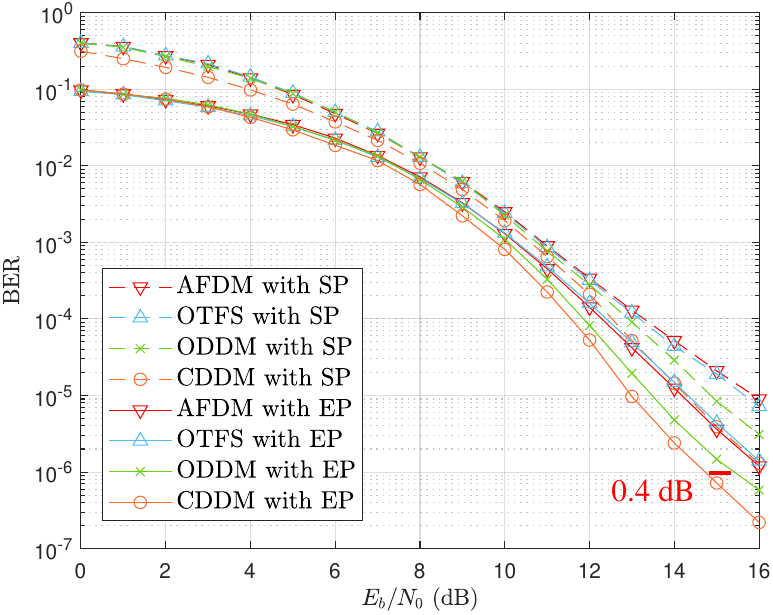}}
	\caption{BER comparison for different pilot schemes at (a) 500 km/h and (b) 120 km/h.}
	\label{CDDMBERCE}
\end{figure}

Fig. \ref{NMSE} presents the NMSE performance of the proposed CDDM channel estimation scheme in comparison with AFDM, OTFS and ODDM. To ensure Doppler interval between pilots using DAFT-based DD chirps, the DD domain dimensions are set as $M_D = N_D = 128$. CDDM, OTFS, and ODDM employ EP with guard intervals \cite{8671740}, and AFDM uses EP in \cite{9562168}. Superimposed pilots scheme (SP) for AFDM uses the scheme in \cite{10711268}, while OTFS and ODDM use the SP scheme in \cite{10333624}. In the SP power setting, CDDM uses $ \sigma_p = 0.3$ and $\lambda = M_D$, while SP in AFDM, OTFS and ODDM occupies 30\% of the total transmit power. The EP power is set to $SNR_p = 60$ dB \cite{8671740} for all waveforms, where $SNR_p$ denotes the pilot SNR. Channel estimation for CDDM-SP is based on the ICZT, while AFDM-SP, OTFS-SP and ODDM-SP adopt the orthogonal matching pursuit (OMP) algorithm. For fairness, all EP-based estimations are performed using threshold method. Although the SP in AFDM from \cite{10711268} can also quickly estimate the shifts using threshold, its performance is relatively poor due to the limited number of pilots. 
In the SP case, the sparse distribution of DAFT-based DD chirps in CDDM enables fast estimation of delay and Doppler shifts via threshold. Moreover, CDDM leverages the correlation of DD chirps for channel estimation, achieving superior performance compared to the OMP-based approaches under high mobility channels. In the EP scheme, both CDDM and ODDM, benefiting from orthogonal pulses and multicarrier modulation in the DD domain, demonstrate superior performance compared to OTFS. Given the same pilot power, the EP in CDDM and ODDM are nearly identical, resulting in almost equivalent channel estimation performance.

Fig. \ref{CDDMBERCE} compares the BER performance under imperfect channel estimation for AFDM, OTFS, ODDM, and CDDM at high-mobility scenarios of 500 km/h and 120 km/h. Both EP and SP are employed for  AFDM, OTFS and ODDM. The configuration of pilots is the same as that for Fig. \ref{NMSE}. For data recovery, MMSE algorithm is adopted by AFDM, OTFS, and ODDM. For the proposed CDDM, MMSE algorithm followed by ICZT is employed. As shown in Fig. \ref{CDDMBERCE}, the EP scheme with guard intervals demonstrates improved performance. In the SP scheme, CDDM outperforms by exploiting the chirp correlation. 

\section{Conclusion} 	\label{Conclusion}
This paper is devoted to a novel 6G waveform, called CDDM, to support future high-mobility communications. With the aid of CZT, we have proposed to spread the data symbols onto the DD chirps by exploiting the resilient chirp signaling properties and the quasi-static feature of the DD domain in high mobility channels. Our study has shown that the proposed CDDM waveform 1) exhibits superior OOBE performance after passing through the SRRC pulse shaping; 2) enables efficient pilot schemes based on the correlation of DD chirps; 3) leads to excellent transmission robustness in terms of the BER and NMSE performances. Although the introduction of CZT requires additional complexity, this is affordable as one can carry out pre-computation for waveform generation.
Finally, it is pointed out that the proposed CDDM waveform holds significant potential for new ISAC development by studying the chirp signaling aided sensing. Therefore, further investigation into ISAC-enabled CDDM constitutes a promising direction for our future research.

\bibliography{referdoc}

\end{document}